\documentclass[aip,jcp,reprint,superscriptaddress,footinbib,floatfix]{revtex4-2}

\usepackage{graphicx}
\usepackage{bm}
\usepackage{amssymb}
\usepackage{amsbsy}
\usepackage{amsmath}
\usepackage{mathtools}
\usepackage{xcolor}
\usepackage{stmaryrd}
\usepackage{mathrsfs}
\usepackage{tikz}
\usetikzlibrary{shapes}
\usepackage{cancel}
\usepackage[normalem]{ulem}
\usepackage{hyperref}
\usepackage{multirow}

\makeatletter
\let\cat@comma@active\@empty
\makeatother

\makeatletter 
\gdef\@ptsize{2}
\let\@currsize\normalsize 
\makeatother 

\usepackage{graphicx}
\usepackage{dcolumn}
\usepackage{bm}

\usepackage[utf8]{inputenc}
\usepackage[T1]{fontenc}
\usepackage{mathptmx}
\usepackage{etoolbox}
\usepackage[shortlabels]{enumitem}
\makeatletter
\def\@email#1#2{%
 \endgroup
 \patchcmd{\titleblock@produce}
  {\frontmatter@RRAPformat}
  {\frontmatter@RRAPformat{\produce@RRAP{*#1\href{mailto:#2}{#2}}}\frontmatter@RRAPformat}
  {}{}
}%

\makeatother

\begin{document}


\title[Linear Viscoelasticity of Semiflexible Polymers]{Linear Viscoelasticity of Semiflexible Polymers with Hydrodynamic Interactions}
\author{Amit Varakhedkar}
 \affiliation{IITB-Monash Research Academy, Indian Institute of Technology Bombay, Mumbai, 400076, India}
 
\author{P. Sunthar}%
\affiliation{ 
Department of Chemical Engineering, Indian Institute of Technology Bombay, Mumbai, 400076, India
}%

\author{J. Ravi Prakash}
 \email{ravi.jagadeeshan@monash.edu}
 \homepage{https://users.monash.edu.au/~rprakash/}
\affiliation{%
Department of Chemical and Biological Engineering, Monash University, 
Melbourne, VIC~3800, Australia
}%

\date{\today}

\begin{abstract}
The linear viscoelastic response of single semiflexible polymer chains in the infinite-dilution limit is studied using Brownian dynamics simulations of coarse-grained bead-spring chains. The springs obey the FENE-Fraenkel force law, a bending potential is used to capture chain stiffness and hydrodynamic interactions are included through the Rotne-Prager-Yamakawa tensor. By calculating the relaxation modulus following a step strain, we demonstrate that the bead-spring chain behaves like an inextensible semiflexible rod over a wide time window with an appropriate choice of spring stiffness and chain extensibility. In the absence of hydrodynamic interactions, our results agree with the existing theoretical predictions for the linear viscoelastic response of free-draining, inextensible, semiflexible rods in the limit of infinite dilution. It is shown that at intermediate times, the stress relaxation modulus exhibits power law behaviour, with the exponent ranging from $(-1/2)$ for flexible chains to $(-5/4)$ for highly rigid chains. At long times, rigid chains undergo orientational relaxation, while flexible chains exhibit Rouse relaxation. Hydrodynamic interactions are found to effect the behaviour at intermediate and long times, with the difference from free-draining behaviour increasing with increasing chain flexibility. Computations of the frequency dependence of loss and storage moduli are found to be in good agreement with experimental data for a wide variety of systems involving semiflexible polymers of varying stiffness across a broad frequency range.

\end{abstract}

\maketitle

\section{Introduction}
Many significant biopolymers are best described as wormlike chains with a persistence length \(l_p\) of the order of the contour length \(L\). Examples include collagen, F-actin, hyaluronic acid, etc. \citep{Alberts2007}. Many of these semiflexible biopolymers form the cytoskeleton of cells, which governs the mechanical rigidity, motility, and adhesion of living cells \citep{Alberts2007}. The mechanical properties of these biopolymers significantly affect the transmission and balance of forces in cells \citep{Jung2020}. Therefore, understanding the mechanical properties of these biopolymers is essential for uncovering the intrinsic mechanisms that underlie cellular functions. One of the key mechanical properties of biopolymers is their linear viscoelasticity, which is quantified by the frequency dependent storage and loss moduli, $G^{\prime}$ and $G^{\prime \prime}$. The moduli typically exhibit a complex dependence on the frequency, displaying various power law regimes at intermediate and high frequencies \citep{Schnurr1997, Koenderink2006}, which arise from the coexistence of different dynamics occurring over different timescales. This paper focuses on the linear viscoelastic response of semiflexible chains in a dilute solution, which serves as a basis from which one can develop an understanding of the behavior of semiflexible polymer solutions at finite concentration, and when they form networks. A mesoscopic model for a dilute solution of semiflexible polymers is introduced here that will enable the understanding of their linear viscoelastic response and the role of bending stiffness and hydrodynamic interactions in determining the observed behavior.

\begin{figure*}[tbph]
\centering
\begin{tabular}{cc}
\includegraphics[width=8cm]{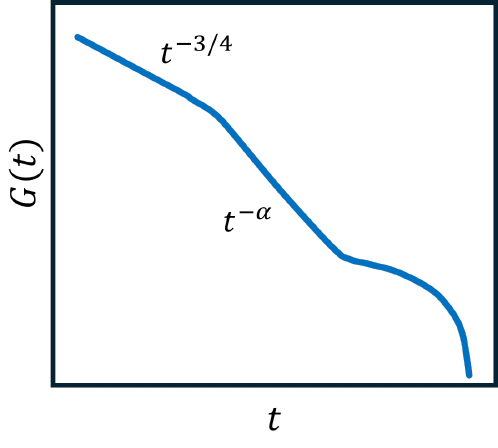} &
\includegraphics[width=8.25cm]{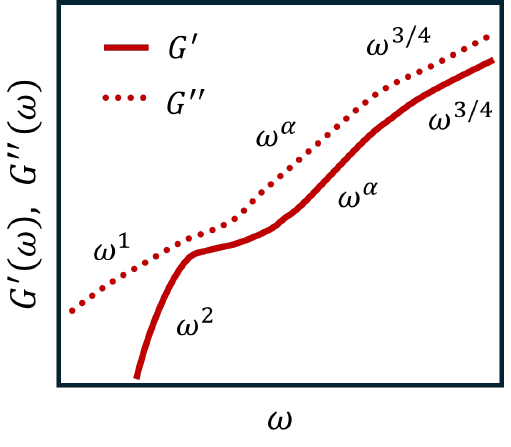} \\
(a) & (b)
\end{tabular}
\caption{Schematic representation of the linear viscoelastic response of a dilute solution of semiflexible polymers. 
(a) Relaxation modulus $G(t)$ as a function of time $t$ following a step strain. 
(b) Storage and Loss moduli, $G^{\prime}(\omega)$ and $G^{\prime \prime}(\omega)$, as a function of frequency $\omega$ obtained by Fourier transforming $G(t)$. 
The exponent $\alpha$ characterizes the power law scaling behavior at the intermediate time and frequency regimes of the relaxation modulus and the dynamic moduli, respectively.}
\label{fig1}
\end{figure*}

Biopolymers are typically broadly classified into four subcategories based on their bending rigidity expressed as the ratio \(L/l_p\) namely, flexible for \(L/l_p \to \infty \), semiflexible when \(L/l_p \approx 1\), stiff semiflexible for \(L/l_p \ll 1\), and rigid rod when \(L/l_p \to 0\). These definitions, however, are strictly relevant only in the dilute or single-chain limit, where the two primary length scales of interest are the contour length \(L\) and the persistence length \(l_p\).  In crosslinked or entangled systems, the characterization of semiflexibility becomes considerably more complex, since additional length scales such as the distance between crosslinks or the entanglement length determine the effective rigidity of the chain over these intermediate length scales.

While the linear viscoelastic response of both flexible and rigid rodlike polymers in dilute solutions is well understood \citep{Doi1988}, a comprehensive quantitative theory for semiflexible polymers over a wide range of bending stiffness is still elusive. Over the years, the understanding of the linear viscoelastic response of dilute solutions of polymers has evolved by studying two limiting cases: rigid rods and flexible chains. A general theory describing the Brownian motion of polymer chains with a fixed bond angle and lengths was first introduced by \citet{Kirkwood1949}. Kirkwood's rigid rod theory \citep{Kirkwood1951}, subsequently considered chains with uniformly spaced, inextensible segments aligned along a straight axis. \citet{Ullman1969} extended this theory to include the effect of the diameter of the rods. In contrast, for flexible chains, \citet{Rouse1953} developed a bead-spring model to capture their viscoelastic properties, which was later refined by \citet{Zimm1956} by including pre-averaged hydrodynamic interactions. Following these early theoretical developments, there have been many studies of bead-rod and bead-spring chain models that have incorporated a number of non-linear microscopic phenomena, \citep{Petera1999, Liu2004, Hur2000, Jendrejack2002, Schroeder2004, Larson2005, Prabhakar2004, Sunthar2005, Saadat2016,Prakash2019, Pincus2023} which are able to capture many of the observed features of these solutions.

Morse and coworkers \citep{Morse1998a, Morse1998b, Morse1998c, Pasquali2001, Shankar2002} developed a comprehensive theoretical framework for semiflexible polymers, extending Kirkwood's rigid-rod theory to account for bending modes and transverse fluctuations along the contour of an inextensible filament, covering regimes from dilute to tightly entangled solutions. The specific formulation of the semiflexible rod theory (SRT) for a single inextensible semiflexible rod in the infinitely dilute limit was presented in detail by \citet{Pasquali2001} and  \citet{Shankar2002}. The SRT particularly applies to inextensible, stiff semiflexible polymers under $\theta$-conditions within the free draining approximation, since it neglects both hydrodynamic interactions and excluded volume effects. The relaxation modulus $G(t)$ predicted by the SRT exhibits three distinct regimes as illustrated  Fig.~\ref{fig1}~(a). At short times, $G(t)$ decays as an exponent of $(-3/4)$ in dilute solutions \citep{Shankar2002}. Interestingly, the $(3/4)$ slope has been observed in dynamic moduli in both theoretical and experimental results for entangled systems \citep{Morse1998c} and crosslinked networks \citep{Gittes1998, Koenderink2006} of semiflexible polymers at high frequencies (which corresponds to short times, as shown schematically in Fig.~\ref{fig1}~(b)). In dilute solutions, this is followed by an intermediate time regime, where the $G(t)$ decays with the exponent $\alpha = 5/4$, due to the sequential relaxation of bending modes \citep{Pasquali2001, Shankar2002}. At long times, after all the intermediate modes have relaxed, the polymer behaves as a rigid rod undergoing orientational relaxation, resulting in a plateau followed by a mono-exponential decay of $G(t)$. As the system transitions towards the semiflexible regime with \(L/l_p \approx 1\), the intermediate power law regime of $(-5/4)$ tends to disappear for more flexible chains, and the initial $(-3/4)$ regime persists over longer timescales before the chain relaxation occurs. These theoretical predictions have been validated by numerical simulations of bead-rod chains \citep{Dimitrakopoulos2001, Shankar2002}, which confirm the time-dependent behavior of the relaxation modulus for stiff semiflexible polymers. The dependence of the dynamic moduli on frequency, which can be obtained by Fourier transformation of $G(t)$ predicted by the SRT also shows excellent agreement with experimental data for poly-$\gamma$-benzyl-L-glutamate (PBLG) polymers \citep{Warren1973, Ookubo1976}, especially for systems with \(L/l_p < 1\), across the entire accessible frequency range. Thus, the SRT provides an excellent benchmark for stiff semiflexible chains ($L/l_p < 1$) with which one can test the validity of a new model, as is done here in the subsequent sections for the present mesoscopic model.

The SRT was developed for inextensible rods with bending modes and assumes that once all internal bending modes are relaxed, the chain behaves as a rigid rod of equivalent length. However, as the chains become more flexible (\( L/l_p > 1 \)), they coil up due to thermal fluctuations rather than rotating as rigid entities. The relaxation modulus obtained from numerical simulations of bead-rod chains by \citet{Shankar2002} revealed noticeable discrepancies with the predictions of the SRT as the chains approach the semiflexible limit of \(L/l_p \approx 1,\) since at large times, the numerical simulations of bead-rod chains do not predict the rod like relaxation predicted by the SRT. \citet{Dimitrakopoulos2001} have also simulated bead-rod chains with Brownian dynamics simulations without hydrodynamic interactions. Rather than identifying multiple power law regimes in $G(t)$ between the plateau region at short times and the behavior at long times, they fit the decay with a single power law at intermediate times. The slope of the power law is shown by them to be varying from $(-5/4)$ for stiff semiflexible chains $(L/lp < 1)$ to $(-1/2)$ for flexible chains.

Most previous studies on semiflexible rods have adapted the free-draining approximation, by neglecting hydrodynamic interactions \citep{Shankar2002, Pasquali2001, Dimitrakopoulos2001}. However, it is well known that hydrodynamic interactions play a crucial role in accurately capturing the dynamic properties of polymers, particularly in dilute solutions \citep{Prakash2019}. For instance, in the case of a rigid rod, incorporating hydrodynamic interactions modifies the prediction of the intrinsic viscosity and terminal relaxation of the rod \citep{Batchelor1970, Kirkwood1951}. Similarly, the linear viscoelastic response of bead-rod chains approaches Rouse behavior without hydrodynamic interactions and Zimm behavior when hydrodynamic interactions are included in a pre-averaged manner \citep{Fixman1974a, Fixman1974b}. For flexible polymers, the experimentally observed linear viscoelastic response agrees excellently with the predictions of the Zimm model \citep{Johnson1970}. However, incorporating hydrodynamic interactions in bead-rod chain models poses a significant challenge. Bead-rod chains require an inextensibility constraint along the contour, which makes it challenging to model their dynamics. Further coupling these with hydrodynamic interactions significantly increases the complexity of simulations. \citep{Ottinger1994, Agarwal1998, Lyulin1999, Petera1999, Agarwal2000, Neelov2002,Liu2004} Although some studies have investigated non-linear rheology in bead-rod chains that include both excluded volume and hydrodynamic interactions \citep{Petera1999, Liu2004}, to our knowledge, there appears to be no models that provide the complete linear viscoelastic response of bead-rod chains with hydrodynamic interactions.

Due to the computational complexity of bead-rod simulations, many modeling efforts have shifted toward coarse-grained bead-spring chain representations. For flexible polymers, a simple bead-spring chain model with Hookean springs is sufficient to reproduce the equilibrium static and dynamic properties \citep{Doi1988, Rubinstein2003, Bird1987}. The first application of this model to semiflexible polymers was introduced by Hearst and coworkers \citep{Harris1966, Hearst1966}, where a semiflexible bead-spring chain formulation predicted a high frequency slope of $(1/4)$ for the dynamic moduli of chains with \(L/l_p = (1/8)\). This result deviates significantly from the $(5/4)$ and $(3/4)$ power law regimes observed in the SRT \citep{Shankar2002}.

Winkler and coworkers \citep{Winkler1994a, Winkler1994b, Harnau1995, Harnau1996, Winkler1997} have carried out seminal work on dilute solutions of semiflexible chains, extending the Rouse model to incorporate bending stiffness while preserving the Gaussian distribution of springs for each chain segment. Their analytical model agrees with the predictions of static properties by the Kratky-Porod wormlike chain model, such as the mean square end-to-end distance and the radius of gyration. \citep{Winkler1994a, Winkler1994b} By incorporating hydrodynamic interactions, the model also investigates various dynamic behaviors, including relaxation times and the dynamic structure factor. However, the treatment of hydrodynamic interactions in their analytical model relies on preaveraging the Rotne-Prager-Yamakawa tensor. Additionally, since the model represents chains using Gaussian springs, inextensibility constraints are not imposed. \citet{Liverpool2001} incorporated an inextensibility constraint along with hydrodynamic interactions using the spatially-resolved Oseen tensor in their model for a semiflexible chain. Their model reports differences in incoherent dynamic light scattering predictions at both low and high times compared to Winkler and coworkers \citep{Harnau1996} and provides a better fit with experimental light scattering data.

More recently, Donev and coworkers \citep{Maxian2023, Maxian2024} developed a sophisticated model where semiflexible chains are modelled as  spectrally coarse-grained inextensible filaments, incorporating fluctuating-hydrodynamic interactions via an integral formulation. This model significantly decreases the computational cost compared to traditional bead-rod models in capturing long time dynamics. While this model represents the most advanced model for modeling inextensible semiflexible chains to date, its application to linear viscoelastic response has not yet been explored.

As mentioned earlier, while bead-rod Brownian dynamics simulations can capture rodlike behavior, including hydrodynamic interactions in these models leads to significant computational challenges. On the other hand, bead-spring chain models are more computationally tractable, but do not accurately reproduce the linear viscoelastic response of semiflexible polymers at high frequencies since they do not account for the inextensibility constraint. The present work aims to develop a mesoscopic model that can reproduce the linear viscoelastic response of semiflexible bead-rod chains and flexible bead-spring chains in dilute solutions, while incorporating hydrodynamic interactions. Although FENE (Finitely Extensible Nonlinear Elastic) spring force laws are commonly used in bead-spring models, they are physically unrealistic for representing short, rigid, and inextensible segments of chains. A few spring force laws developed with a non-zero rest length, such as the Fraenkel spring force law \citep{Fraenkel1952} approach the limit of an inextensible rod at higher spring stiffness values. However, the Fraenkel spring force law is a linear spring force law; it requires extremely large spring stiffness values to enforce the inextensibility limit. A more versatile alternative is the FENE-Fraenkel spring force law, which can behave as both a rod and an entropic spring and transition between extreme cases \citep{Hsieh2006, Pincus2020, Pincus2023}. This force law has an additional extensibility parameter, allowing the spring to approach the extensibility limit by appropriately tuning the extensibility and spring stiffness parameters. It allows the replication of bead-rod chain behavior without the added complexities of explicitly imposing inextensibility in the Brownian dynamics simulations. Recent studies have shown that FENE-Fraenkel springs can replicate the behavior of bead-rod chains in shear and extensional flows at a reduced computational cost \citep{Pincus2023}. In this work, a single semiflexible chain in the limit of infinite dilution is modeled as a bead-spring chain, with the beads connected by FENE-Fraenkel springs. Appropriate spring parameters are identified for which the bead-spring chains can replicate the linear viscoelastic response of a bead-rod chain. A bending potential is introduced along the chains to tune the persistence length. This model is validated against existing theoretical predictons, numerical simulations, and experimental studies by comparing their results with the predictions of both the relaxation modulus, $G(t)$ and dynamic moduli, $G^{\prime}$ and $G^{\prime \prime}$. Additionally this study examines the influence of fluctuating hydrodynamic interactions on the linear viscoelastic response of semiflexible chains with a range of bending stiffness for the first time.

The rest of the paper is structured as follows. Sec.~\ref{sec:gov_eqn} introduces the bead-spring chain model for polymers, presents the governing equations and various intramolecular interactions including the bending potential and spring force laws. The nondimensionalising units adopted in this study and conversion formulae for different unit systems are presented in Sec.~\ref{sec:diff_units}. This is followed by the description of the simulation parameters and various viscoelastic properties computed in this work in Sec.~\ref{sec:param} and Sec.~\ref{sec:properties}, respectively. The validity of the proposed model to predict dynamic properties for semiflexible chains with hydrodynamic interactions is demonstrated in Sec.~\ref{sec:valid} by showing agreement with the results of multi-particle collision dynamics (MPCD) simulations of \citet{Nikoubashman2016}. Sec.~\ref{sec:spr_selection} presents relaxation modulus for stiff semiflexible chains with different spring force laws to establish the FENE-Fraenkel spring chains as the ideal choice. The optimum spring parameters and number of beads required to get exact agreement with the SRT and existing bead-rod simulation results \citep{Shankar2002} are determined in Sec.~\ref{sec:free_drain}. The relaxation modulus for semiflexible chains with varying bending stiffness with and without hydrodynamic interactions are presented in Sec.~\ref{sec:Llp_effect} and Sec.~\ref{sec:hyd_intr} respectively. Sec.~\ref{sec:hyd_intr} also presents the intermediate slopes calculated from the relaxation modulus for different chains to highlight the influence of hydrodynamic interactions on chains with bending stiffness. In Sec.~\ref{sec:exp}, the present model is compared with the experimental data for the linear viscoelasticity response of PBLG \citep{Warren1973} and Collagen \citep{Nestler1983} in infinitely dilute solutions as well as the SRT. Sec.~\ref{sec:conclusion} summarizes the key conclusions and outlines directions for future work.

\section{\label{sec:methods} Model formulation}

\subsection{\label{sec:gov_eqn} Governing equations}
A coarse grained bead-spring chain model is considered to simulate semiflexible polymer solutions using Brownian dynamics. The position vector $\bm{r}_{\mu} (t)$ of each bead $\mu$ is evolved in time $t$ using a first-order Euler integration scheme, which numerically solves the It\^o stochastic differential equation governing its Brownian motion \citep{Ottinger2012}
\begin{align}\label{eq:bd}
\bm{r}_{\mu}(t + \Delta t) = \ & \bm{r}_{\mu}(t) + \frac{\Delta t}{4} \sum_{\nu=1}^{N_b} [\bm{D}_{\mu \nu} \cdot \bm{F}_{\nu}] + \frac{1}{\sqrt{2}} \sum_{\nu=1}^{N_b} [\bm{B}_{\mu \nu} \cdot \Delta \bm{W}_{\nu}].
\end{align}
The equation is nondimensionalised using Hookean units with length scale $l_H=\sqrt{{k_{\text{B}}T}/{\hat H}}$ and time scale $\lambda_H={\zeta}/{4{\hat H}}$, where $\hat H$ is the spring stiffness and $\zeta=6\pi \eta_{\text{s}} a$ is the Stokes friction coefficient of the bead of radius $a$ and solvent viscosity $\eta_{\text{s}}$. Here, $N_b$ is the number of beads per chain. $\Delta \bm{W}_{\nu}$ is the nondimensional Wiener process that accounts for the stochasticity in the above equation. The components of $\Delta \bm{W}_{\nu}$ are obtained from real values of a Gaussian distribution that has zero mean and variance as \(\Delta t\). $\bm{B}_{\mu \nu}$ is a nondimensional tensor which is evaluated by the decomposition of the diffusion tensor \(\bm{D}_{\mu \nu}\) defined as:
\begin{equation}
\label{eq:7_3}
\bm D_{\mu\nu} = \delta_{\mu\nu} \bm{\delta} + \bm{\Omega}_{\mu\nu}
\end{equation}
where $\delta_{\mu \nu}$ is the Kronecker delta, $\bm{\delta}$ is a unit vector, and \(\bm{\Omega}_{\mu \nu}\) is the hydrodynamic interaction tensor. The matrices $\mathcal{D}$ and $\mathcal{B}$ are defined as block matrices with \(N \times N\) blocks, each having dimensions of $3 \times 3$ such that the $(\mu,\nu)$-th block of $\mathcal{D}$ contains the components of the diffusion tensor $\bm{D}_{\mu\nu }$, whereas, the corresponding block of $\mathcal{B}$ is equal to $\bm{B}_{ \mu\nu}$. The decomposition rule for obtaining $\mathcal{B}$ is expressed as $\mathcal{B} . \mathcal{B}^{T} = \mathcal{D}$. 

In the present study, the Rotne-Prager-Yamakava (RPY) tensor is used to compute hydrodynamic interactions.
\begin{equation}
\label{eq:7_9}
\bm{\Omega}_{\mu \nu} = \bm{\Omega} (\bm{r}_{\mu} - \bm{r}_{\nu})
\end{equation}
where,
\begin{equation}
\bm{\Omega} (\bm{r}) = \Omega_{1} \bm{\delta} + \Omega_{2} \frac{\bm{r}\bm{r}}{r^{2}}
\end{equation}
The terms \(\Omega_{1}\) and \(\Omega_{2}\) correspond to first and second-order corrections due to hydrodynamic interactions. These are evaluated as,
\begin{equation*}
\Omega_1 = \begin{cases} \dfrac{3\sqrt{\pi}}{4} \dfrac{h^*}{r}\left({1+\dfrac{2\pi}{3}\dfrac{{h^*}^2}{{r}^{2}}}\right) & \text{for} \quad r\ge2\sqrt{\pi}h^* \\
 1- \dfrac{9}{32} \dfrac{r}{h^*\sqrt{\pi}} & \text{for} \quad r\leq 2\sqrt{\pi}h^* 
\end{cases}
\end{equation*}
\vspace{5pt}
\begin{equation*}
\Omega_2 = \begin{cases} \dfrac{3\sqrt{\pi}}{4} \dfrac{h^*}{r} \left({1-\dfrac{2\pi}{3}\dfrac{{h^*}^2}{{r}^{2}}}\right) & \text{for} \quad r\ge2\sqrt{\pi}h^* \\
 \dfrac{3}{32} \dfrac{r}{h^*\sqrt{\pi}} & \text{for} \quad r\leq 2\sqrt{\pi}h^* 
\end{cases}
\end{equation*}
The hydrodynamic interaction parameter \( h^* \) is the dimensionless bead radius defined as \(h^* = a / \sqrt{\pi k_{B} T /{\hat H}}\). 

\begin{figure}[bh]
\centering
\includegraphics*[width=8.5cm]{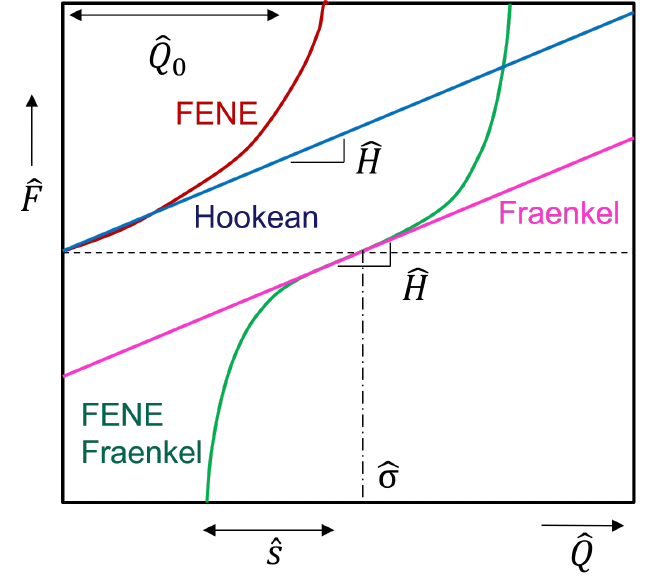}
\caption{Force versus extension curves for different spring laws: Hookean (Blue), FENE (Red), Fraenkel (Magenta) and FENE-Fraenkel (Green). \(Q_{0}\) denotes the FENE spring stretchability parameter, \(\sigma\) is the Fraenkel spring's natural length and \(s\) is the spring extensibility around \(\sigma\)}
\label{fig2}
\end{figure}

The net nondimensional force $\bm{F}_{\nu}$ acting along the bead ${\nu}$ is a sum of the spring, bending and excluded volume interaction forces:
\begin{equation}
\bm{F}_{\nu} = \bm{F}_{\nu}^{S} + \bm{F}_{\nu}^{SDK} + \bm{F}_{\nu}^{B}
\end{equation}
The spring force $\bm{F}_{\nu}^{S}$ is taken to be a Finitely Extensible Nonlinear Elastic - Fraenkel (FENE-Fraenkel) spring force law. This force law can simultaneously represent other force laws, such as the Hookean, FENE and Fraenkel springs (Fig.~\ref{fig2}) by an appropriate choice of parameter values. The FENE-Fraenkel spring force law written in dimensional form is,
\begin{align}\label{FF_force_eqn}
\hat{\bm{F}^{S}} = \frac{\hat{H}(\hat{Q}-\hat{\sigma})}{1-(\hat{Q}-\hat{\sigma})^2/(\hat{s})^2} \frac{\hat{\bm{Q}}}{\hat{Q}}    
\end{align}
where dimensional quantities are denoted by $(\hat{\ })$. $\hat{\bm{Q}}$ corresponds to the bead vector with length $\hat{Q}$, $\hat{\sigma}$ is the natural rest length of the spring, and $\hat{H}$ is the spring stiffness with the units of force per unit length. The parameter $\hat{s}$ corresponds to the stretchability of the spring along the length $\hat{\sigma}$. The spring can approach the limit of a rod of equivalent length $\hat{\sigma}$ for extremely high values of $\hat{H}$. Alternatively, the spring also mimics an inextensible rod for small values of stretchability $\hat{s}$. As can be seen in Fig.~\ref{fig2}, setting $\hat{\sigma}=0$ in \eqref{FF_force_eqn} recovers the FENE spring, where $\hat{s}$ is equivalent to the extensibility parameter $\hat Q_0$ typically used to represent extensibility in FENE springs. In the limit of $\hat{s} \rightarrow \infty$, the Hookean spring force law is recovered. On the other hand, a finite value of $\hat{\sigma}$ in the limit of $\hat{s} \rightarrow \infty$, leads to the Fraenkel spring force law. In nondimensional form, when rescaled in terms of Hookean units, the spring force law reads as:
\begin{align}\label{FF_force_eqn_Hk}
{\bm{F}^{S}} = \frac{(Q - \sigma)}{1 - (Q - \sigma)^2 / s^2}
\frac{\bm{Q}}{Q}   
\end{align}

The force $\bm{F}_{\nu}^{SDK}$ corresponds to the forces due to the excluded volume interactions between beads. These interactions are modeled using a piecewise potential proposed by the Soddemann, Duenweg and Kramer (SDK) \citep{Soddemann2001}:
{\small
\begin{align}\label{eq:SDK} 
\frac{\hat{U}_{\mu\nu}^{\text{SDK}}}{k_{\text{B}} T} = \left\{
\begin{array}{l l}
4 \! \left[ \! \left( \dfrac{d}{  r_{\mu\nu}} \right)^{12} - \left( \dfrac{d}{  r_{\mu\nu}} \right)^6 + \dfrac{1}{4} \right] - \epsilon \, ;& r_{\mu\nu}\leq 2^{1/6}d \\ [20pt]
\dfrac{1}{2}\,  \epsilon \! \left[ \cos \! \left(\! \alpha \left( \dfrac{ r_{\mu\nu}}{d} \right)^2+ \beta \! \right) - 1 \right] \hspace{-1pt} ; & \!\!  \hspace{-15pt} 2^{1/6}d \leq  r_{\mu\nu} \leq r_{\rm c} \\ [20pt]
0 &   r_{\mu\nu} \geq  r_{\rm c}
\end{array}\right.
\end{align}
}
This potential acts between any two interacting beads $\mu$ and $\nu$, separated by a distance $r_{\mu\nu}= \left\vert\bm{r}_{\mu} - \bm{r}_{\nu}\right\vert$. Here $\epsilon$ is the attractive well depth, and $\alpha$ and $\beta$ are the parameters where the potential smoothly goes to zero at a cutoff radius $r_c = 1.82 d$ \citep{Santra2019}. The value of nondimensional distance $d$ is set equal to $1$. Interpolation between a good solvent, $\theta$-solvent and a poor solvent can be obtained by a suitable choice of \(\epsilon\). When $\epsilon=0$, the SDK potential models good solvents, similar to the Weeks-Chandler-Andersen (WCA) potential \citep{Andersen1971}. The SDK potential is used in the present study to validate the results of \citet{Nikoubashman2016} in Sec.~\ref{sec:valid}, who have studied chains in athermal solvents. Apart from Sec.~\ref{sec:valid}, the effect of excluded volume interactions is not studied in the present work, since most of the studies of semiflexible chains \citep{Shankar2002, Dimitrakopoulos2001} with which our results are compared with are in theta solvents.

\begin{figure}[th]
\centering
\includegraphics*[width=8.5cm]{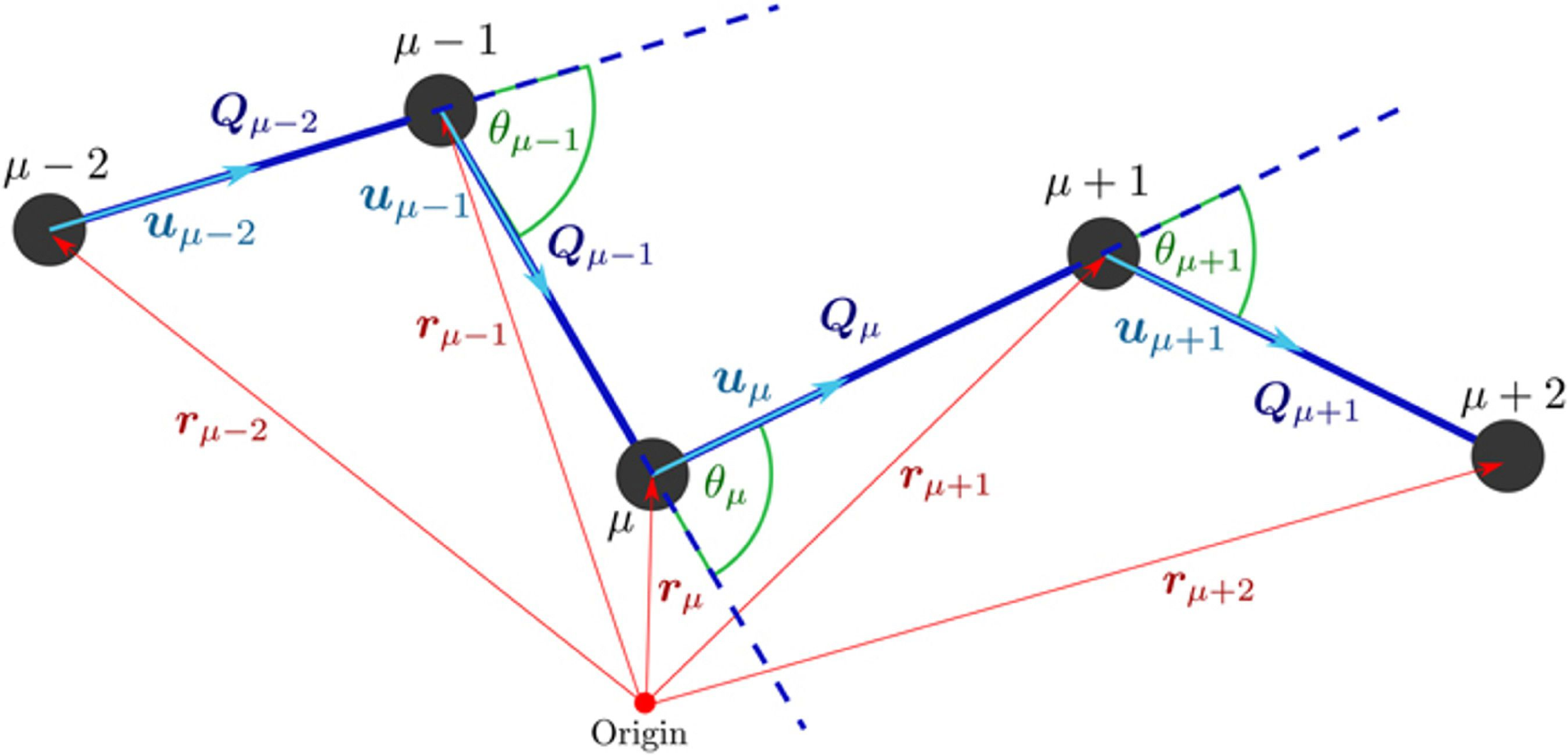} 
\caption{Diagram illustrating the labeling scheme for beads, segments, and included angles. The position $\bm{r}_{\mu}$ of bead $\mu$ relative to the center of mass. $\bm{u}_{\mu}$ is the unit vector of the segment connecting beads $\mu$ and $\mu+1$ with length $Q_{\mu}$. Included angle $\theta_{\mu}$ is the angle between unit vectors $\bm{u}_{\mu}$ and $\bm{u}_{\mu-1}$. This figure has been reproduced from \citet{Pincus2023} with permission.}
\label{fig3}
\end{figure}

Semiflexibility is an important characteristic of a biopolymer, which in the bead-spring chain model is captured using a potential that imposes an energetic penalty based on the angle $\theta_{\mu}$ between successive bond vectors as illustrated in Fig.~\ref{fig3}. The bending potential is given by,
\begin{align}\label{eq:7_6}
\displaystyle \hat{U}_{\mu}^{B} = C k_{B} T(1-\cos{\theta_{\mu}})
\end{align}
where \(\hat U_{\mu}^{B}\) and \(\theta_{\mu}\) are the bending potential and included angle between the vectors ${\bm{\hat Q}}_{\mu}$ and ${\bm{\hat Q}}_{\mu+1}$. $C$ is the bending stiffness of the chain. The force on a bead \(\mu\) due to the bending potential \citep{Pincus2023} is given by,
\begin{align}
\label{eq:bend}
\frac{\hat{\bm{F}}_\mu^{(\mathrm{b})}}{k_B T} = C \Bigg\{ 
& \left[ 
\frac{1}{\hat Q_\mu} \left( \bm{u}_\mu \cos\theta_\mu - \bm{u}_{\mu-1} \right) \right. \nonumber \\
& + \left. \frac{1}{\hat Q_{\mu-1}} \left( -\bm{u}_{\mu-1} \cos\theta_\mu + \bm{u}_\mu \right) 
\right] \nonumber \\
& + \left[
\frac{1}{\hat Q_{\mu-1}} \left( -\bm{u}_{\mu-1} \cos\theta_{\mu-1} + \bm{u}_{\mu-2} \right) 
\right] \nonumber \\
& + \left[
\frac{1}{\hat Q_\mu} \left( \bm{u}_\mu \cos\theta_{\mu+1} - \bm{u}_{\mu+1} \right)
\right] 
\Bigg\}
\end{align}
Here, $\bm{u}_{\mu}$ is the unit vector from bead $\mu$ to $\mu+1$ with dimensional length $\hat Q_{\mu}$. A useful relationship between the bending stiffness $C$ and $L / l_{p}$ is given by \citet{Saadat2016}, where the number of Kuhn steps $N_{k,s}$ in each segment of the chain is a function of $L / l_{p}$,
\begin{align}\label{eq:Nks}
N_{k,s} = \frac {L} {2 (N_{b}-1) l_{p}}
\end{align}
\begin{align}\label{eq:7_7}
\displaystyle C = \frac {1 + p_{b,1}(2N_{k,s}) + p_{b,2}(2N_{k,s}^{2})} {2N_{k,s} + p_{b,3}(2N_{k,s}^{2}) + p_{b,4}(2N_{k,s}^{3})}
\end{align}
Here, the $p_{b,i}$ = 1.237, 0.8105, 1.0243, 0.4595 for $i = 1,
2, 3, 4$ respectively. This is a Padé approximation chosen to exactly match the nearest-neighbor correlation of a continuous wormlike chain \citep{Saadat2016}.
This relationship enables the definition of semiflexibility in terms of the $L / l_{p}$.

\subsection{\label{sec:diff_units} Nondimensionalization}

Since we would like to compare our results with existing results for the SRT and bead-rod simulations, it is also possible to use another system for nondimensionalising the simulation results, which we denote as \textit{rodlike} units.
\begin{align}\label{eq:8_2}
l_R \equiv \hat \sigma, \quad \lambda_R \equiv \frac{\zeta {\hat \sigma}^2}{k_B T}, \quad F_R \equiv \frac{k_B T}{\hat \sigma}
\end{align}
This system is normally used for bead-rod chains with rod length $L$, however in the present study, the rest length $\hat \sigma$ is used instead. For Hookean and FENE springs which do not have a finite rest length, $\hat \sigma$ is set equal to $\langle \hat  Q \rangle$ which is the equilibrium spring length. The nondimensional spring stiffness in rodlike units is denoted as ${H_{R}} = \hat H {\hat \sigma}^2 / k_B T$ \citep{Pincus2020}. This is an important parameter by tuning which, a rodlike response can be replicated with springs, as shown in Sec.~\ref{sec:free_drain}. Time can also be normalised with respect to $\lambda_{Rod}$, which corresponds to the terminal relaxation of a discretized rigid rod of \(N_b\) beads \citep{Bird1987} with each bead spaced at a distance $\hat \sigma$ apart, given by:
\begin{equation}
\label{eq:rod}
 \lambda_{Rod} = \frac{\zeta {\hat \sigma}^2 (N_{b}^{3} - N_{b})}{72 k_B T}
\end{equation}
It is straightforward to convert between Hookean and rodlike units. For example, the length and time in rodlike units can be obtained via,
\begin{align*}
Q_{R} &= \frac{\hat{Q}}{l_R} =  \left( \frac{l_H}{l_R} \right) Q =  \left( \frac{ \sqrt{k_BT}}{\hat \sigma \sqrt{\hat H}} \right) Q ; \\ 
t_{R} & = \frac{\hat{t}}{\lambda_R} = \left( \frac{\lambda_H}{\lambda_R} \right) t  = \left( \frac{k_BT}{4 \hat H{\hat \sigma}^2} \right) t \nonumber
\label{eq:8_combined}
\end{align*}

\subsection{\label{sec:param} Simulation parameters}

The beads are connected by FENE-Fraenkel springs with finite stiffness, such that the average bond length corresponds to the rest length \(\hat \sigma\). In this paper, the spring parameters are set as $\hat \sigma = 3 l_H$ and $\hat s = 2 l_H$ unless stated otherwise. As a result, the contour length of the chain is given by \(L = (N_b - 1) \hat \sigma\). The simulations were performed either in the free draining approximations, without hydrodynamic interactions \((h^* = 0)\), or with hydrodynamic interactions \((h^* = 0.2)\). All the simulations were conducted in the infinitely dilute regime.

The spring stiffness parameter \({H_{R}}\) is varied over several orders of magnitude, ranging from \(10^{2}\) to \(10^{7}\). The nondimensional time step was chosen to be \(\Delta {t_{R}} = 1 \times 10^{-3}\) for the simplest case with \({H_{R}} = 100\). To accurately capture the dynamics of stiff springs, $\Delta {t_{R}}$ was progressively reduced with increase in \({H_{R}}\), for example, decreasing to $\Delta {t_{R}} = 1 \times 10^{-4}$ for $H_{R} = 1000$, $\Delta {t_{R}} = 1 \times 10^{-5}$ for ${H_{R}} = 10000$, and so on. The time step convergence for each $H_{R}$ is ensured. This reduction is necessary since the smallest relaxation time of springs scales inversely with \({H_{R}}\) \citep{Fraenkel1952}.

The semiflexibility of the chains is varied from the flexible limit, with \(L/l_p \to \infty\) to the stiff semiflexible limit of $L/l_p = 0.125$. For each chain, the rotational relaxation time $\lambda_{Rod}$ of a rod of equivalent contour length was calculated. Each independent trajectory consisted of an initial equilibration phase lasting $3 \times \lambda_{Rod}$, followed by a production phase of $7 \times \lambda_{Rod}$. The dynamic properties were calculated as a function of time in the production phase for individual trajectories, followed by calculating the ensemble averages and error estimate over 30,000 independent trajectories. Typical parameter values used in the simulations are listed in Table~\ref{parametervalues}.

    \begin{table*}[t]
\normalsize
\caption{Typical parameter values used in the Brownian dynamics simulations}
\label{parametervalues}
\renewcommand{\arraystretch}{1.3}
\begin{tabular*}{\textwidth}{@{\extracolsep{\fill}}llllll}
  \hline
  {} & Parameter & Symbol & Values & & \\
  \hline
  1 & Number of beads per chain         & ${N}_{b}$     & 8, 16, 24, 32 & & \\
  2 & Rodlike spring stiffness          & ${H}_{R}$     & $10^{2}$ to $10^{7}$ & & \\
  3 & Hydrodynamic interaction parameter& $h^{*}$           & 0, 0.2 & & \\
  4 & Integration time step             & $\Delta t_{R}$& $10^{-3}$ to $10^{-8}$ & & \\
  5 & Semiflexibility                   & $L / l_p$     & 0.125 to $\infty$ & & \\
  \hline
\end{tabular*}
\end{table*}

\subsection{\label{sec:properties} Estimating linear viscoelastic properties}

In this section we will discuss the procedure to calculate various linear viscoelastic properties from the Brownian dynamics simulations.

\subsubsection{Stress Tensor}
The dynamic properties such as relaxation modulus, zero-shear rate viscosity and dynamic moduli investigated in this work can be defined in terms of the components of the stress-tensor for the polymer solution. The nondimensional contribution to the stress tensor is given by the Kramers-Kirkwood expression,
\begin{equation}\label{Kramers}
\bm{\tau}_{\text{p}} = \frac{1}{N_{c}} \left\langle \sum_{\xi=1}^{N_{c}} \sum_{\nu=1}^{N_b} \left(\bm{r}_\nu^{(\xi)}-\bm{r}_c^{(\xi)} \right) \bm{F}_{\xi\nu} \right\rangle
\end{equation}
where the stress is nondimensionalised by $N_{c} \, (k_B T/V)$, with $N_{c}$ being the number of chains in a solution of volume $V$. In the context of the simulations carried out in this work, all results correspond to an isolated single chain, representing the infinitely dilute limit. In this regime, the number of chains $N_c$ is always equal to 1 in Eq.~\eqref{Kramers} and in all subsequent equations. The force on each bead $\mu$ in a chain $\xi$ is given by,
\begin{equation}\label{totalforce}
\bm{F}_{\xi\nu} = \sum_{\beta=1}^{N_{c}} \sum_{\substack{\mu=1 \\ \mu\ne\nu}}^{N_b} \bm{F}_{\xi\nu,\beta\mu}^{\text{SDK}} + \sum_{\substack{\mu=1 \\ \mu\ne\nu}}^{N_b} \bm{F}_{\xi\nu,\xi\mu}^{\text{S}} + \bm{F}_{\xi\nu}^{\text{B}}
\end{equation}
It is important to note that the bending force, as defined in Eq.~\eqref{eq:bend}, is not a pairwise force but instead involves successive triplets of beads. A dimensionless stress tensor $\bm{S}$ which accounts for the total contribution to the stress tensor from the chain for a single independent run is be defined by,
\begin{equation}\label{S}
\bm{S} = \sum_{\xi=1}^{N_c} \sum_{\nu=1}^{N_b} \left( \bm{r}_\nu^{(\xi)}-\bm{r}_c^{(\xi)} \right) \bm{F}_{\xi\nu}
\end{equation}
which can be simplified to,
\begin{align}
\label{Kramersv2}
\bm{S} &=  
\frac{1}{2} \sum_{\nu=1}^{N}\sum_{\substack{\mu=1 \\ \mu\ne\nu}}^{N} 
   \bm{r}_{\nu\mu} \bm{F}_{\nu\mu}^{\text{SDK}} 
   + \sum_{\xi=1}^{{N}_{c}} \sum_{i=1}^{N_b-1} 
   \bm{Q}_{i}^{(\xi)} \bm{F}^{{\text{S}}}\!\left( \bm{Q}_{i}^{(\xi)} \right) \notag \\
& + C \sum_{\xi=1}^{{N}_{c}} \sum_{i=1}^{N_b-2} 
   \Big[ \bm{u}_i^{(\xi)}\bm{u}_{i+1}^{(\xi)} 
   + \bm{u}_{i+1}^{(\xi)}\bm{u}_i^{(\xi)}  \notag \\
&   - \cos \theta_{i+1}^{(\xi)}\!\left( 
     \bm{u}_i^{(\xi)}\bm{u}_i^{(\xi)} 
   + \bm{u}_{i+1}^{(\xi)}\bm{u}_{i+1}^{(\xi)} \right) \Big].
\end{align}
Once the stress tensor is computed, we can easily estimate various dynamic properties and material functions for the polymer solution. Here, the focus is on calculating linear viscoelastic properties in terms of the relaxation modulus, the zero shear rate viscosity, and dynamic moduli.

\subsubsection{Relaxation Modulus}
The relaxation modulus \(G(t)\) is obtained from equilibrium simulations  using the Green-Kubo relation that relates \(G(t)\) to the autocorrelation function of the stress tensor. At equilibrium, the stress tensor is isotropic, hence the relaxation modulus is given by the expression,
\begin{equation}
G(t) = \frac{1}{3} \left( G_{xy}(t) + G_{xz}(t) + G_{yz}(t) \right)
\end{equation}
where the components \(G_{ij}(t)\) (which are equal to each other at equilibrium) are given by the expression,
\begin{equation}
G_{ij}(t) = \frac{1}{N_c} \left\langle S_{ij}(0) \, S_{ij}(t) \right\rangle
\end{equation}
 The relaxation modulus can be easily computed using the above equation.
\begin{figure}[b]
\centering
\includegraphics*[width=8.5cm]{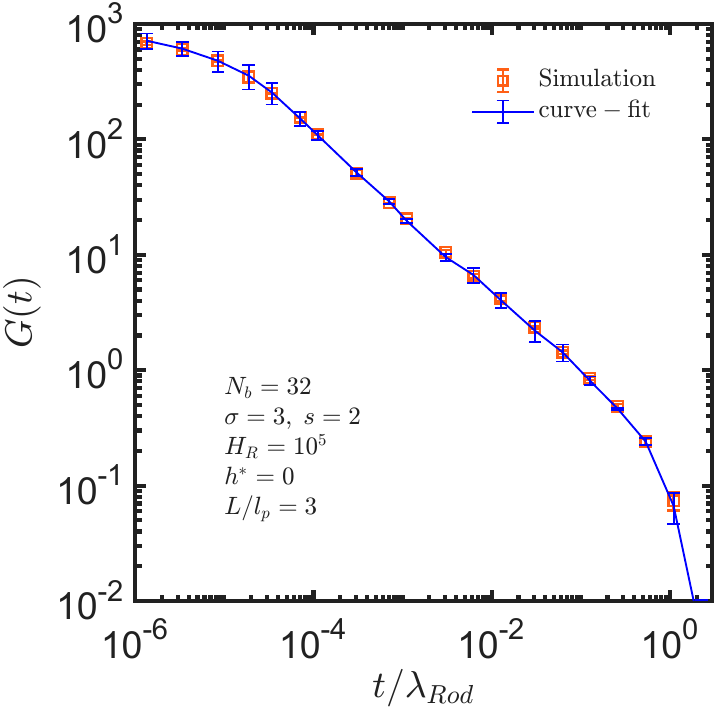}
\caption{The nondimensional relaxation modulus $G(t)$ as a function of scaled time for $N_b=32$ FENE--Fraenkel chain with $H_R = 1 \times 10^{5}$ and $L/l_p = 3$. 
The blue curve is a fit to the simulation data using a sum of exponential functions with 7 exponents.}
\label{fig4}
\end{figure}
As shown in Fig.~\ref{fig4} for a typical example, the relaxation modulus obtained from the simulations is fitted with a sum of exponentials,
\begin{equation}
G(t) = \sum_{i=1}^{n} a_i \exp(- b_i t)
\end{equation}
where $a_i$ and $b_i$ are the fitting parameters and $n$ is the number of exponentials required to fit the curve. All the relaxation modulus curves evaluated in this paper are fitted using $5$ to $9$ exponentials. The errorbars of the fitted $G(t)$ is obtained from the envelope of curves reconstructed using the upper and lower bounds of the coefficients. All the subsequent calculations involving the relaxation modulus are carried out using similar fits.

\subsubsection{Zero Shear Rate Viscosity}
While the study of shear viscosity at moderately high shear rates \(\dot{\gamma}\) is important in non-linear rheology, the study of linear viscoelasticity primarily focuses on the polymeric component of the zero-shear rate viscosity, defined as: $ \eta_{p,0} = \lim_{\dot{\gamma} \to 0} \eta_p$. Here \(\eta_{p,0} \) is calculated from equilibrium simulations by integrating the relaxation modulus $G(t)$ \citep{Fixman1981, Lee2009},
\begin{equation}
\eta_{p,0} = \int_0^\infty G(t) \, dt
\end{equation}

\begin{figure*}[tbph]
\centering
\begin{tabular}{cc}
\includegraphics[width=8.5cm]{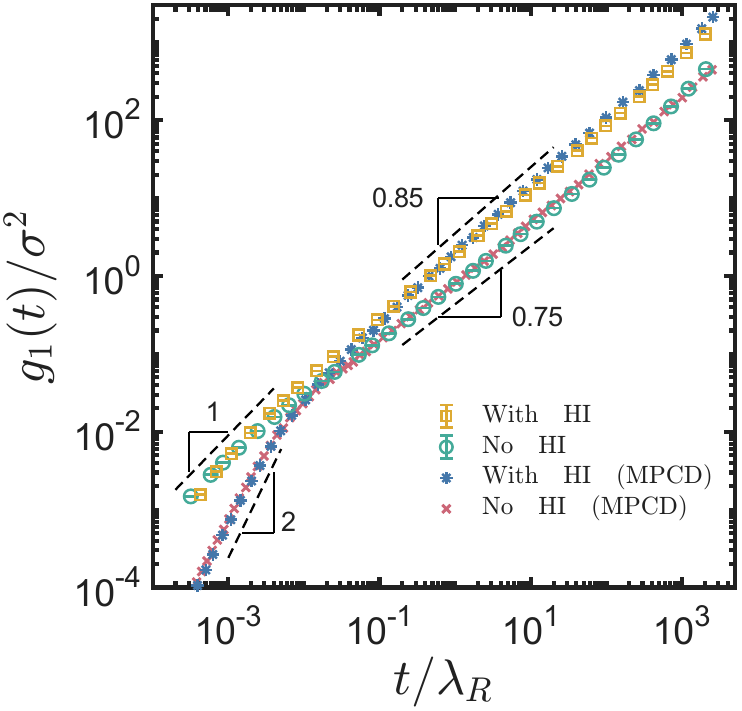} &
\includegraphics[width=8.5cm]{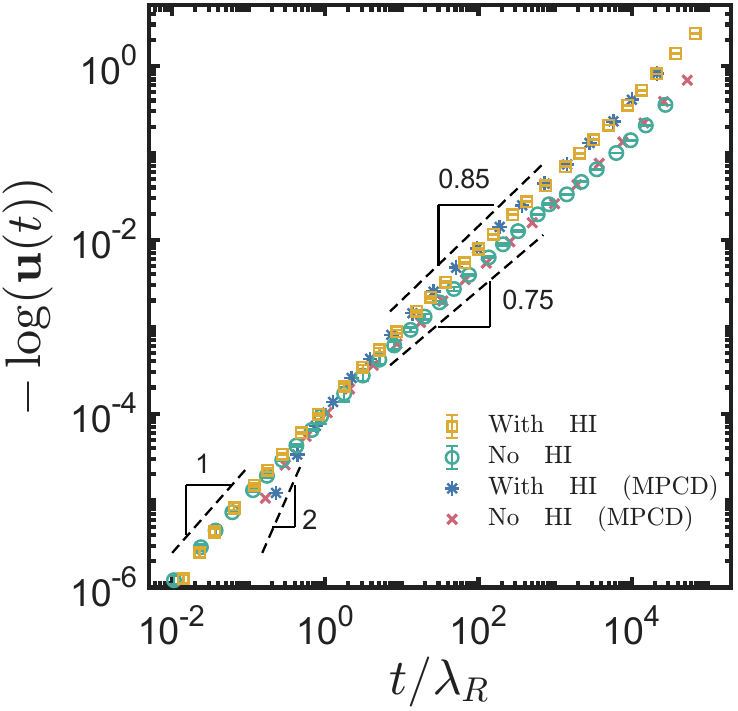} \\
(a) & (b)
\end{tabular}
\caption{(a) Internal monomer mean square displacement \(g_{1}(t)\) and (b) end-to-end unit vector autocorrelation function \(\bm{u}(t)\) for a semiflexible chain with bending stiffness \(C=20\) and \(N_{b}=48\) beads. with HI (squares) and without HI (circles). The results of \citet{Nikoubashman2016}, with and without HI, are denoted by asterisks (*) and crosses ($\times$), respectively.}
\label{fig5}
\end{figure*}

\subsubsection{Dynamic Moduli}
The elastic and viscous response of a viscoelastic fluid is generally characterised by the storage \(G'\) and the loss \(G''\) moduli, together referred to as the dynamic moduli. These properties are typically obtained from oscillatory shear flow experiments. In the current simulations, in the limit of very small strain amplitude, \(G'\) and \(G''\) are determined from a Fourier transformation of the relaxation modulus $G(t)$ \citep{Wittmer2015}, 
\begin{equation}
G'(\omega) = \int_{0}^{\infty} G(t) \sin(\omega t) \, d(\omega t)
\end{equation}
\begin{equation}
G''(\omega) = \int_{0}^{\infty} G(t) \cos(\omega t) \, d(\omega t)
\end{equation} 

\section{Results}

\subsection{\label{sec:valid} Code validation}

To validate the implementation of hydrodynamic interactions in the simulation code, the results are compared with those of \citet{Nikoubashman2016} who used multi particle collision dynamics (MPCD) simulations with hydrodynamic interactions captured through the presence of explicit solvent molecules to examine the dynamics of semiflexible chains. Since, to our knowledge there appears to be no previous work on the linear viscoelastic response of a semiflexible chain with fluctuating hydrodynamic interactions, the validation focuses on a different set of dynamic properties examined by \citet{Nikoubashman2016} for this system. The key properties examined here are the inner monomer mean square displacement \(g_1(t)\) and the end-to-end unit vector autocorrelation function \(\bm{u}(t)\). An ensemble of FENE spring chains is simulated using the same parameters as \citet{Nikoubashman2016}, namely, a chain length of \(N_b = 48\) with FENE extensibility parameter \(Q_0 = 1.5\), spring stiffness of \(H = 30\) and bending stiffness \(C = 20\). Results are presented for an athermal solvent. Since the FENE spring does not have a rest length, the equilibrium spring length $\langle Q \rangle$ is taken as equivalent to $\sigma$ such that the results can be presented in rodlike units. It is important to note that, while the MPCD algorithm models solvent particles explicitly, Brownian dynamics simulations accounts for hydrodynamic interactions implicitly via nontrivial long-range dynamic correlations in the stochastic displacements. The results without hydrodynamic interactions reported in \citep{Nikoubashman2016} were obtained using Brownian dynamics with an implicit solvent representation.
The mean square displacement of an inner monomer, \(g_{1}(t)\), is defined as:
\begin{equation}
g_1(t) = \langle [\bm{r}_\mu(t) - \bm{r}_\mu(0)]^2 \rangle
\end{equation}
where \(\mu\) denotes an inner monomer away from the chain ends. Since we use a 48-bead chain, we compute \(g_{1}(t)\) at \(\mu = 24\).
The end-to-end unit vector ($R_e$) autocorrelation function, \(\bm{u}(t)\), is given by:
\begin{equation}
\bm{u}(t) = \langle {R_e}(t) \cdot {R_e}(0) \rangle
\end{equation}
Fig.~\ref{fig5}~(a) illustrates the mean-square displacement of a tagged bead, \(g_1(t)\), for a semiflexible chain with and without hydrodynamic interactions. In both cases, three characteristic regimes are observed. At short times, MPCD simulations capture the ballistic diffusion regime with a time-scaling exponent of \(2\), which is absent in the Brownian dynamics simulations that transition directly to the intermediate fully damped regime. In this regime, \(g_1(t)\) scales as \(0.75\) without hydrodynamic interactions, whereas the inclusion of hydrodynamic interactions leads to a faster decay that deviates with a power law slope of $0.85$. At long times, both cases should recover the diffusive regime with exponent \(1\), but the onset of diffusive regime for a semiflexible chain with large bending stiffness occurs at considerably longer times, which is not included here. In both cases, Brownian dynamics simulations are in excellent agreement with MPCD.  

Fig.~\ref{fig5}~(b) shows the corresponding end-to-end unit vector autocorrelation function, \(\bm{u}(t)\), for chains with and without hydrodynamic interactions. A similar overall trend is evident: hydrodynamic interactions decays faster compared to the free-draining case. The Brownian dynamics simulations reproduce the MPCD–molecular dynamics results over a wide time window, with deviations only at very short times corresponding to the ballistic regime. Fig.~\ref{fig5}~(a) and (b) validate the accuracy of the present approach in capturing the relevant dynamics of semiflexible chains.

While \citet{Nikoubashman2016}'s study provides valuable insights into the dynamics of stiff Gaussian chains, they did not study the linear viscoelastic behavior of semiflexible polymers with hydrodynamic interactions.

\subsection{\label{sec:free_drain} Free Draining Limit}

One of the key aims of using FENE-Fraenkel springs is to reproduce the linear viscoelastic response of a semiflexible bead-rod chain. This section identifies the appropriate spring parameters for which a FENE-Fraenkel spring chain emulates a bead-rod chain. All results in this section are presented in rodlike units to facilitate comparison with existing data for semiflexible bead-rod chains and theoretical predictions.

\subsubsection{\label{sec:spr_selection} Comparison of different spring force laws}

\begin{figure}[t]
\centering
\includegraphics*[width=8.5cm]{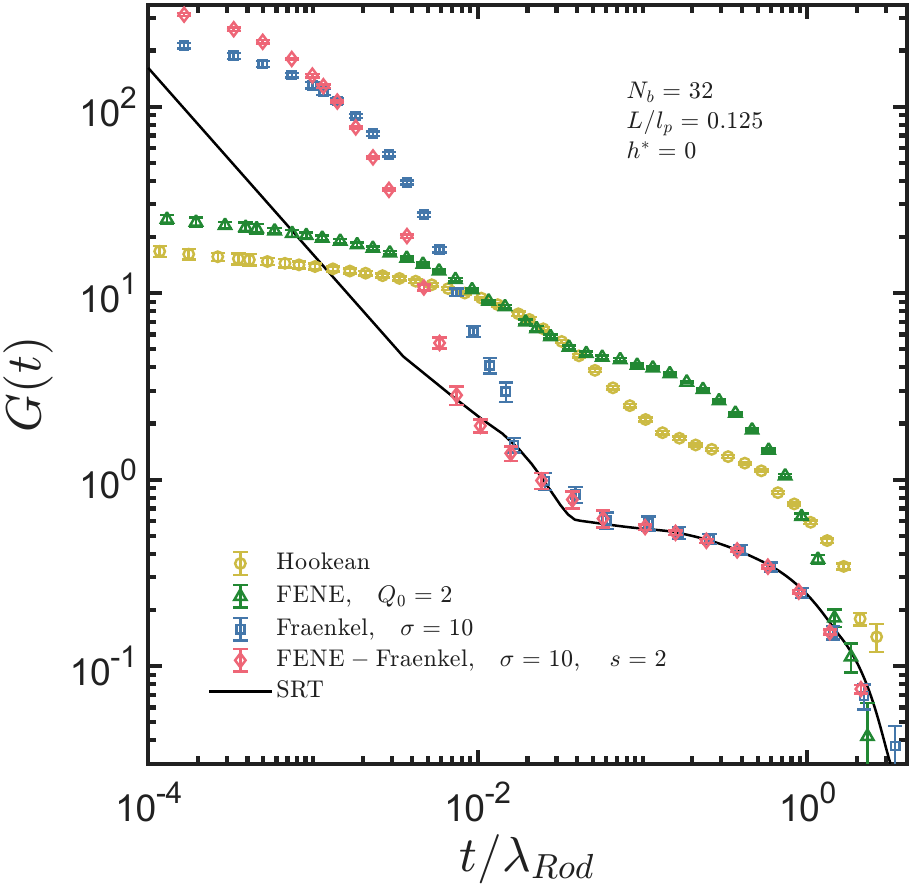}
\caption{Effect of spring forces (a) Hookean, (b) FENE, (c) Fraenkel, and (d) FENE-Fraenkel on the relaxation modulus of a semilexible chain of \(N_{b}\) = 32 with \(L / l_{p} = 0.125\). The black line denotes the analytical expression of \(G(t)\) from the SRT for \(L / l_{p} = 0.125\) \citep{Shankar2002}. Here time is nondimensionalised with respect to rodlike units}
\label{fig6}
\end{figure}

Fig.~\ref{fig6} shows the relaxation modulus for different spring force laws in a 32 bead chain at \(L/lp = 0.125\). The parameters used are $Q_{0}=2$ for FENE springs, $\sigma = 10$ for Fraenkel and FENE-Fraenkel Springs (with $s=2$). The choice of stretchability parameter $s$ was based on the smallest permissible value for the given timestep of the simulation as outlined by \citet{Pincus2020}. The results were nondimensionalised with respect to the terminal relaxation time of a discretized rigid rod \(\lambda_{Rod}\) \citep{Bird1987}. Unlike Fraenkel and FENE-Fraenkel springs, FENE and Hookean springs do not have a finite rest length. For these springs, the equilibrium length of the spring $\langle \hat Q \rangle$ was used instead of \(\hat \sigma\) for calculation of \(\lambda_{Rod}\). The results are compared with the $G(t)$ obtained from the SRT \citep{Shankar2002} for a semiflexible chain of $L/l_p = 0.125$.  

The results demonstrate that springs with a finite rest length, such as Fraenkel and FENE-Fraenkel springs, more accurately capture the long-time behavior of a semiflexible rod of equivalent stiffness. Their relaxation modulus curves exhibit a clear two-stage relaxation process: an initial rapid decay in $G(t)$ followed by a slower relaxation that agrees closely with the relaxation modulus predicted by the SRT. The fast initial decay arises from the relaxation of individual springs in the chain, consistent with the behavior of a Fraenkel dumbbell, where a rapid early relaxation is followed by the dumbbell relaxing as a rigid rod of rest length \citep{Fraenkel1952}.  

\begin{figure}[t]
\centering
\includegraphics*[width=8.5cm]{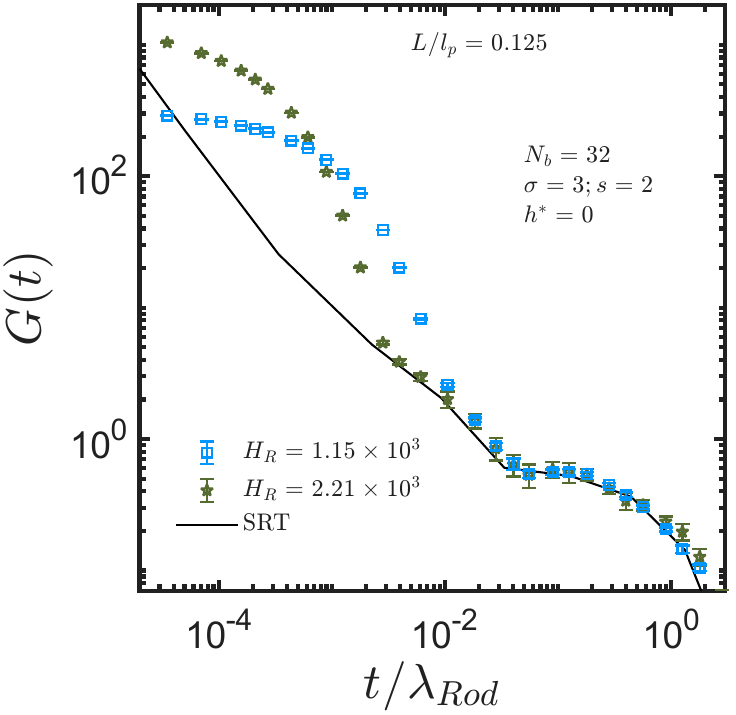}
\caption{The nondimensional relaxation modulus $G(t)$ as a function of scaled time for $N_b=32$ FENE-Fraenkel chain with $H_R = 1.15 \times 10^{3}$ (blue) and $H_R = 2.21 \times 10^{3}$ (green) for $L/l_p = 0.125$. The spring parameters are kept constant with $\sigma = 3$ and $s = 2$. The black curve represents the analytical expression of the SRT \citep{Shankar2002} for $L/l_p = 0.125$.}
\label{fig7}
\end{figure}

Once this initial spring relaxation is complete, both Fraenkel and FENE-Fraenkel spring chains align well with the analytical SRT expression for semiflexible rods \citep{Shankar2002}. They accurately reproduce the relaxation of intermediate modes and capture the terminal plateau associated with the orientational relaxation of a rigid rod. Between the two, the FENE-Fraenkel springs offers a modest improvement: their spring relaxation occurs earlier, extending the time window over which the chain behaves like a bead-rod system. Although the improvement is small, it is achieved without additional computational cost, making FENE-Fraenkel springs a preferred choice for future simulations.  

In contrast, Hookean and FENE springs fail to reproduce the expected terminal plateau (Fig.~\ref{fig6}). Their negligible rest length causes the initial spring relaxations to transition smoothly to the terminal relaxation, thereby suppressing a distinct plateau regime. Moreover, these models deviate from the SRT even at shorter times, failing to capture the intermediate relaxation modes.

\subsubsection{\label{sec:Hr_effect} Effect of increased spring stiffness}

\begin{figure}[t]
\centering
\includegraphics*[width=8.5cm]{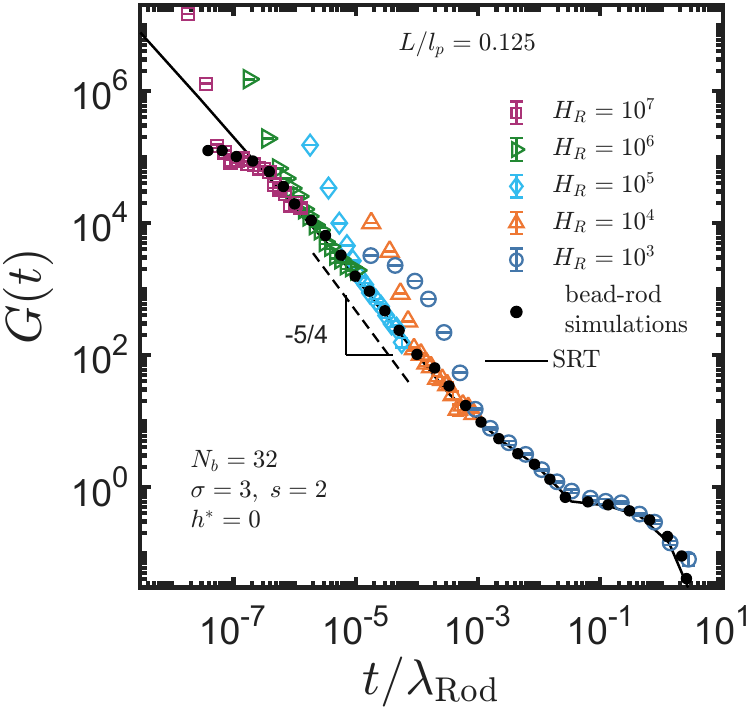}
\caption{Master curve of the nondimensional relaxation modulus $G(t)$ as a function of scaled time for different \(H_{R}\) varying from \(10^{3}\) to \(10^{7}\) from right to left, for a \(N_{b} = 32\) FENE-Fraenkel chain with \(L / l_{p}=0.125\). Results compared with analytical expression of $G(t)$ from the SRT (black curve) and semiflexible bead-rod simulations of \(N_{b}=32\) (black symbols) reproduced from \citet{Shankar2002}.}
\label{fig8}
\end{figure}

As previously discussed in Sec.~\ref{sec:gov_eqn}, the two methods for a FENE-Fraenkel spring used to mimic an inextensible rod involve increasing the spring stiffness and decreasing the stretchability parameter for a constant rest length. This subsection will focus on determining the critical value of the dimensionless spring stiffness, $H_R$, which allows the spring to effectively behave as a rigid rod.

A preliminary comparison of two $H_R$ values with the SRT analytical relaxation modulus (Fig.~\ref{fig7}) reveals a clear trend: higher $H_R$ values lead to earlier completion of spring relaxation, enabling better agreement with the analytical model over a wider time window. In contrast, lower $H_R$ values delay spring relaxation. Importantly, once the springs have relaxed, the curves for both $H_R$ values collapse onto each other, indicating that at sufficiently long times the choice of $H_R$ no longer influences the overall bead–rod chain relaxation. This observation motivated the extension of simulations to larger $H_R$ values to capture the relaxation of intermediate modes at earlier times.

Fig.~\ref{fig8} presents a master curve of the relaxation modulus $G(t)$ for a chain with $N_b = 32$ beads and $L/l_p = 0.125$ across different values of $H_R$. The rest length ($\sigma$) is kept fixed, so $H_R$ depends only on the spring stiffness $H$. Increasing $H_R$ shifts the spring relaxation to earlier times, thereby extending the effective bead–rod regime. Since the curves converge at long times, the data is shown only up to the point where the higher $H_R$ curves merge with those of the lower $H_R$.  

For \(H_R = 10^{7}\), an additional plateau appears at the onset of spring relaxation. This plateau corresponds to the initial plateau of a bead-rod chain with the same number of beads and arises from discretizing a continuous chain into a finite number of beads. Beyond this point, at long times the modulus follows the bead-rod behavior exactly, confirming that if the initial spring relaxations at very short times is ignored, with the appropriate choice of spring parameters, a bead-spring chain can reproduce the linear viscoelastic response of a bead-rod chain over the entire subsequent time range.

A further key observation is the emergence of an intermediate slope of $(-5/4)$, consistent with the SRT predictions for stiff semiflexible rods with $L/l_p = 0.125$ \citep{Shankar2002}. At intermediate times the chains relax rapidly, while at long times they approach the terminal plateau governed by rotational diffusion of an inextensible rod. These results demonstrate that a semiflexible FENE–Fraenkel chain can faithfully capture both the intermediate mode relaxations and the orientational relaxation of an inextensible semiflexible rod.

\begin{figure}[t]
\centering
\includegraphics*[width=8.5cm]{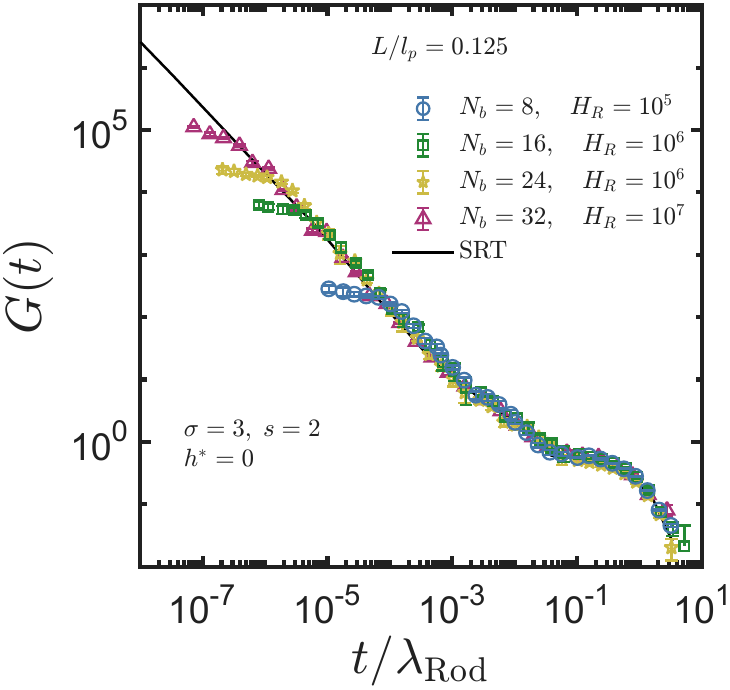}
\caption{The nondimensional relaxation modulus $G(t)$ as a function of scaled time for different \(N_{b}\) varying from \(N_{b} = 8\) to \(32\) from right to left for a FENE-Fraenkel chain with \(L / l_{p}=0.125\). The black curve represents the analytical expression of the SRT \citep{Shankar2002} for $L/l_p = 0.125$. Initial spring relaxation is not displayed in all the cases.}
\label{fig9}
\end{figure}

\subsubsection{\label{sec:Nb_effect} Effect of the number of beads}

To further validate the model, the effect of chain discretization was examined by plotting $G(t)$ for a fixed $L/l_p=0.125$ across different bead numbers (Fig.~\ref{fig9}). Note that the initial spring relaxation before the appearance of the first bead-rod plateau is not displayed to keep the focus on comparison with the SRT prediction of the relaxation modulus. The results demonstrate that, when appropriately scaled, the curves collapse onto a master curve. This indicates a universal behavior that is independent of the level of discretization, consistent with previous findings \citep{Shankar2002, Dimitrakopoulos2001}.

It is important to note how the $G(t)$ curves are constructed for a semiflexible chain with given $N_b$ and $L/l_p$. For each $N_b$, the curve is obtained from a master curve representing various values of the dimensionless spring stiffness $H_R$ (Fig.~\ref{fig8}). The highest $H_R$ value at which the initial bead–rod plateau is observed is reported for each $N_b$. Since the curves for different $H_R$ values collapse onto the master curve after the initial spring relaxation, the $G(t)$ curves are stitched together by neglecting these initial relaxations. This procedure yields a single representative $G(t)$ curve for a discretized semiflexible chain.  

A clear difference emerges between chains with increasing numbers of beads. Chains with lower discretization (e.g., $N_b = 8$) exhibit a bead-rod plateau only at later times and fail to capture the early-time dynamics. In contrast, chains with higher discretization (e.g., $N_b = 32$) develop a bead-rod plateau at earlier times, thereby reproducing the relaxation modulus of a semiflexible rod over a wider time window. This highlights the importance of sufficient discretization to capture a wider dynamic range of a continuous semiflexible chain. Moreover, the collapse of curves for different $N_b$ and $H_R$ values onto a single master curve provides strong evidence that the observed relaxation behavior is an intrinsic property of the semiflexible chain and not an artifact of the simulation parameters.

The observed dependence of the relaxation modulus on the level of discretization can be further rationalized in terms of the scaling of the relaxation times of bending modes. As shown analytically by \citet{Harnau1995} and discussed in the review by \citet{Dreiss2007}, the relaxation rate of bending modes in semiflexible chains follows the scaling relation $\lambda_q \sim q^4$, where $q$ is the bending wavevector associated with the mode. In the stiff semiflexible or rodlike limit ($L/l_p < 1$), this scaling arises from the fourth-order spatial derivative in the governing Langevin equation of the wormlike chain, reflecting bending-controlled dynamics rather than contour fluctuations. \citep{Harnau1995}

In the simulations, the smallest resolvable bending length scale is approximately proportional to $(L/N_b)$, where $L$ is the contour length and $N_b$ is the number of beads. This corresponds to the shortest bending wavelength that can be captured in the simulation. The associated bending wavevector for this mode is the largest accessible one, given by $q_{\text{max}} \sim \pi N_b / L$. Since the relaxation times of these modes scales as $\lambda_q \sim q^4$, doubling $N_b$ effectively doubles $q_{\text{max}}$, thereby extending the time window by roughly a factor of $2^4 = 16$. This scaling signifies that resolving finer bending wavelengths introduces faster relaxation modes, thereby broadening the time window of rodlike relaxation behavior.

The trend observed in Fig.~\ref{fig9} is consistent with this interpretation: as $N_b$ increases, the simulation captures progressively shorter bending modes, and the relaxation modulus $G(t)$ follows the semiflexible rodlike behavior predicted by the SRT over a wider time window before reaching the orientational-relaxation plateau at long times. This clearly demonstrates that the level of chain discretization directly controls the accessible dynamic range of bending relaxation in the simulations.

Despite increasing the discretization, the occurance of the $(-3/4)$ power law at short times remained elusive. Capturing this scaling for stiff semiflexible chains would require simulations with a very large number of beads, which is computationally prohibitive \citep{Shankar2002, Dimitrakopoulos2001}. Nevertheless, the results demonstrate the successful capturing of the relaxation of intermediate modes.

\subsubsection{\label{sec:Llp_effect} Effect of semiflexibility}
\begin{figure}[tbph]
\centerline{
\begin{tabular}{c}
\includegraphics[width=7cm,height=!]{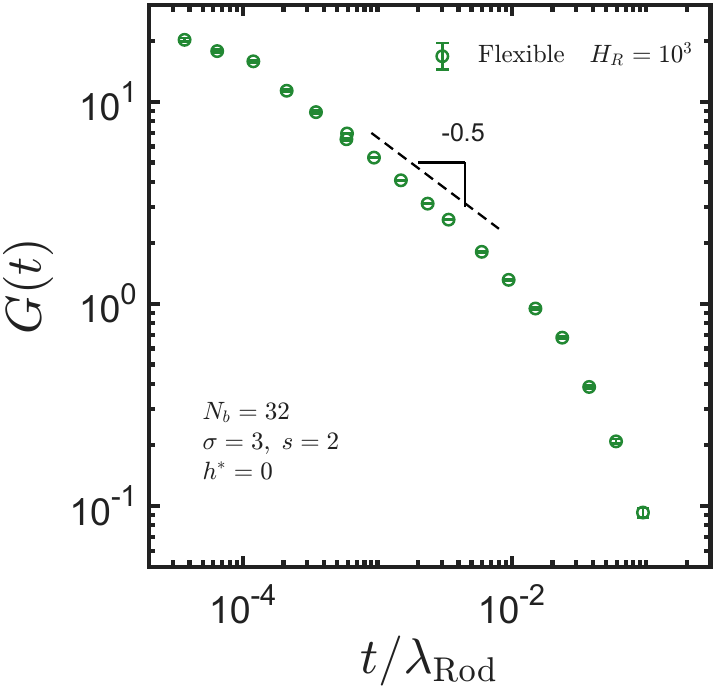} \\
(a) \\[6pt]
\includegraphics[width=7cm,height=!]{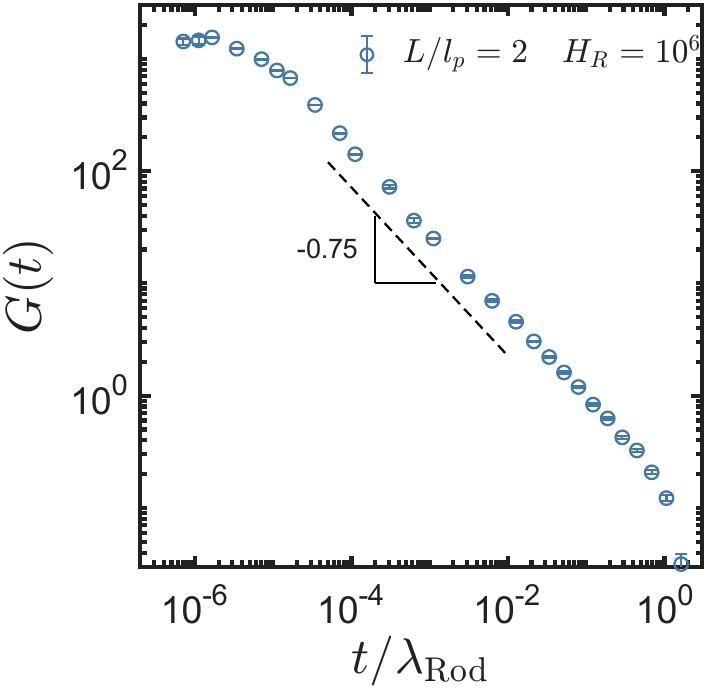} \\
(b) \\[6pt]
\includegraphics[width=7cm,height=!]{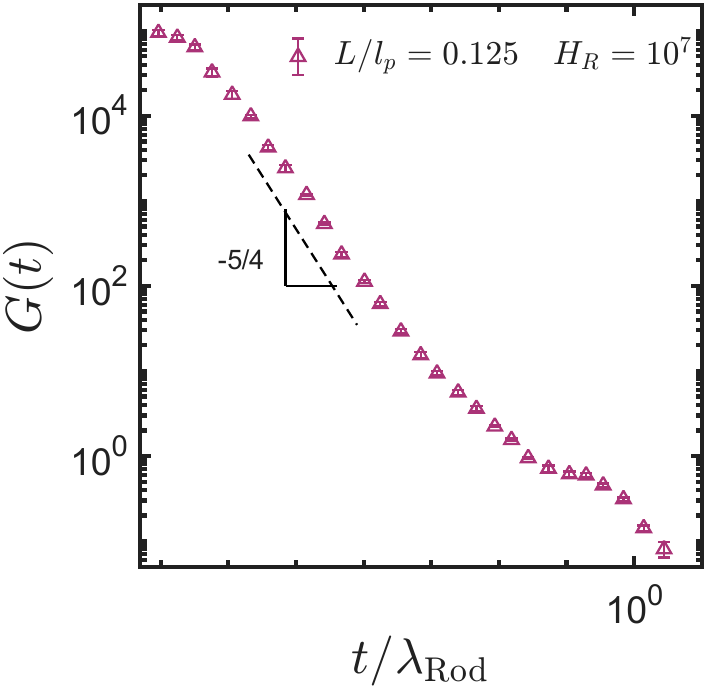}\\
(c) \\
\end{tabular}
}
\vspace{-5pt}
\caption{The nondimensional relaxation modulus $G(t)$ as a function of scaled time with different semiflexibilities, (a) Flexible chain, (b) \(L / l_{p} = 2\) and (c) \(L / l_{p} = 0.125\) is shown for \(N_{b}=32\) FENE-Fraenkel chain. Here the slopes represent the intermediate power law scaling for each cases. Initial spring relaxation is not displayed in all the cases.}
\label{fig10}
\end{figure}

Fig.~\ref{fig10} shows the relaxation modulus \(G(t)\) as a function of time for different values of \(L / l_{p}\), spanning the fully flexible to the stiff semiflexible regime (\(L / l_{p} = 0.125\)). After the initial bead-rod plateau, \(G(t)\) exhibits a distinct power-law scaling before transitioning to long-time behavior. This scaling persists over a wide time window, with the exponent varying from $(-1/2)$ for the fully flexible chain as displayed in Fig.~\ref{fig10}~(a) to $(-5/4)$ for the stiff semiflexible chain with \(L/l_p = 0.125\) (Fig.~\ref{fig10}~(c)). The $(-1/2)$ slope is in excellent agreement with the Rouse model predictions, while the $(-5/4)$ slope matches the predictions of the SRT \citep{Shankar2002, Pasquali2001}. 
 
\begin{figure*}[tbp]
\centering
\begin{tabular}{ccc}
  \includegraphics[width=7cm]{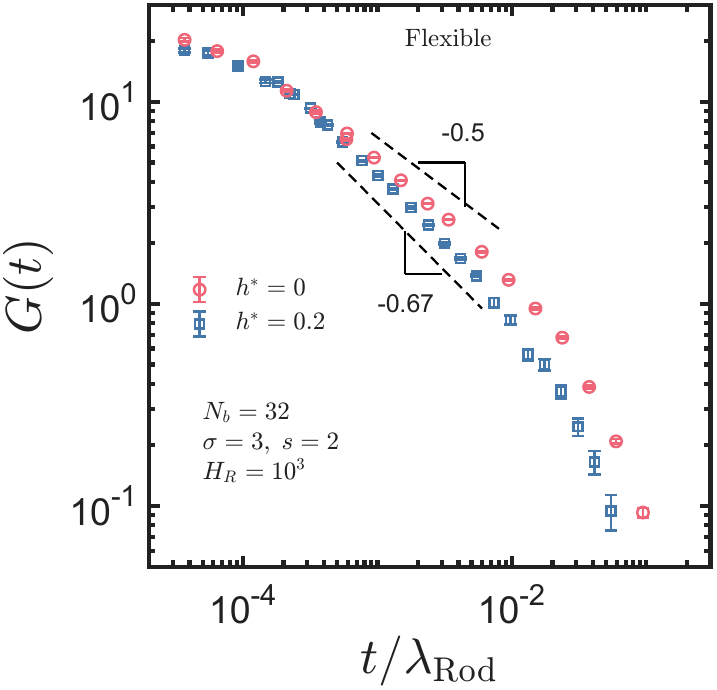} &
  \includegraphics[width=7cm]{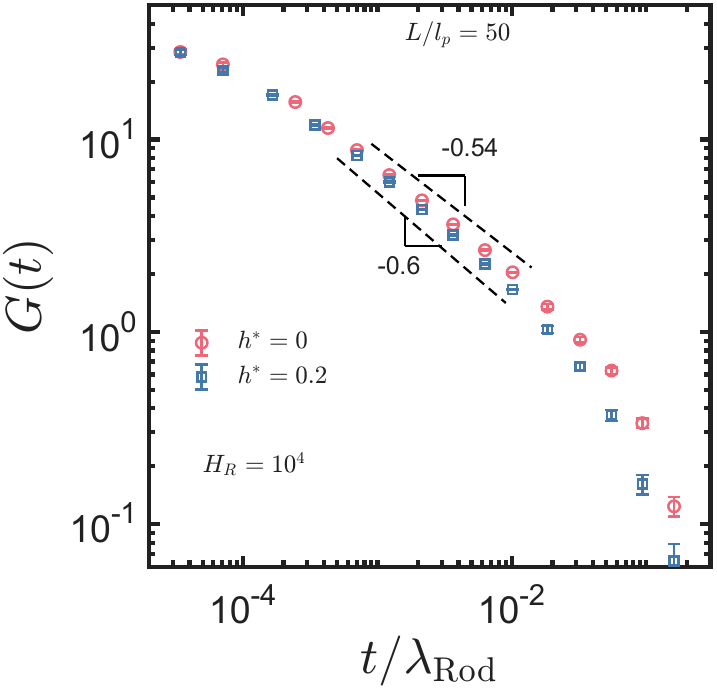} \\
  (a) & (b) \\[10pt]
  \includegraphics[width=7cm]{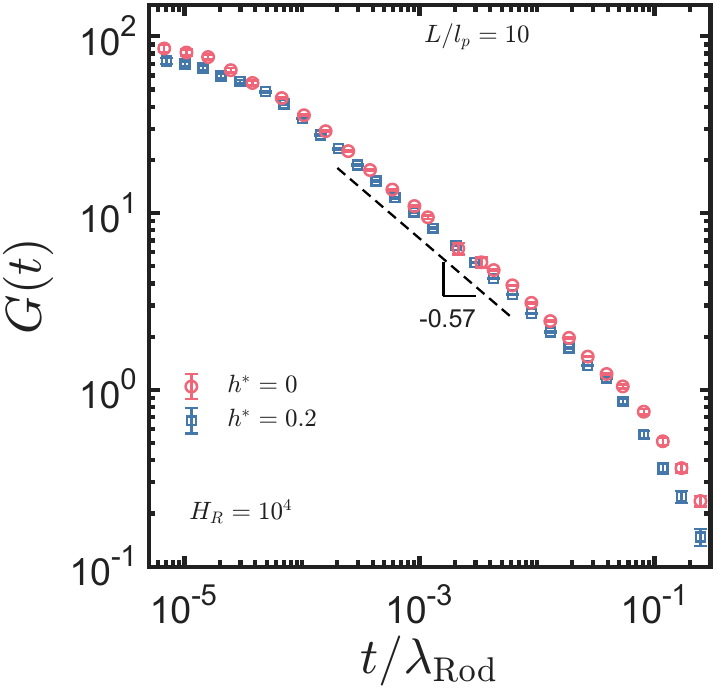} &
  \includegraphics[width=7cm]{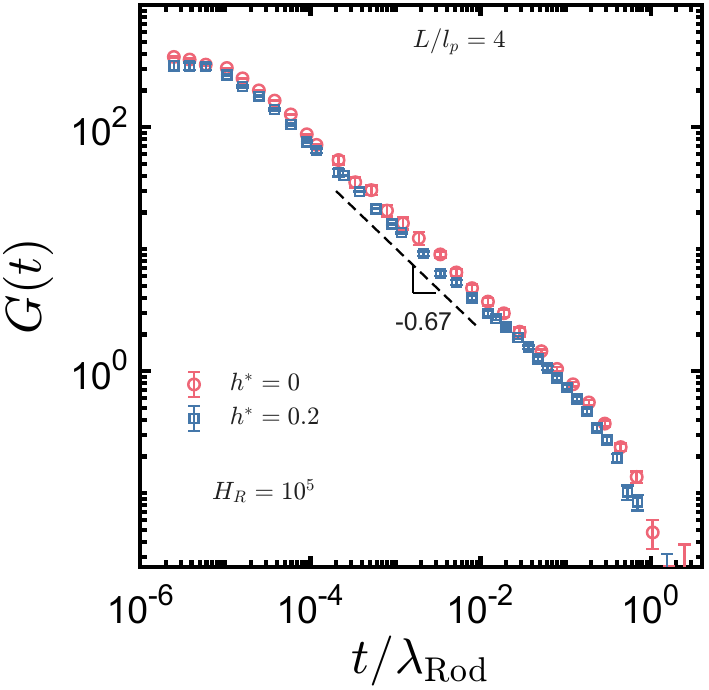} \\
  (c) & (d) \\[10pt]
  \includegraphics[width=7cm]{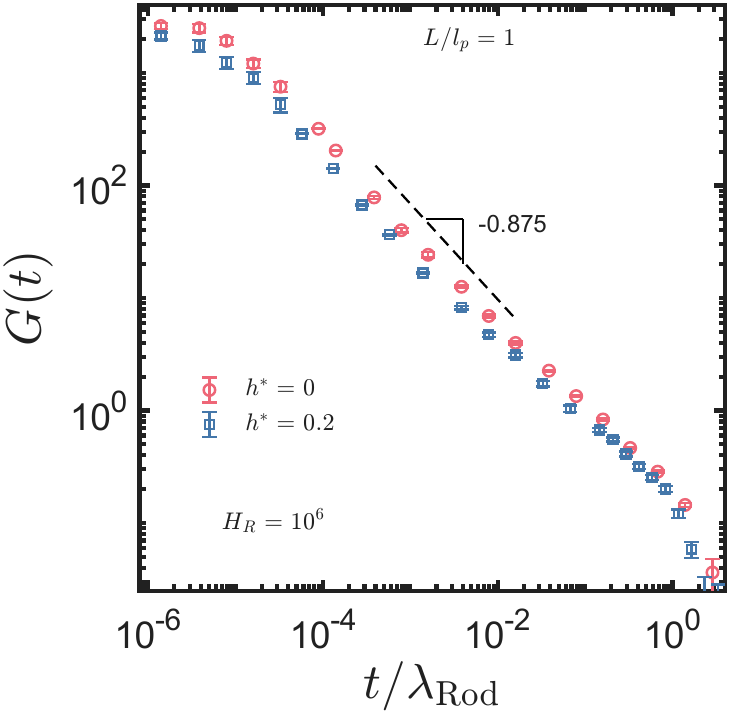} &
  \includegraphics[width=7cm]{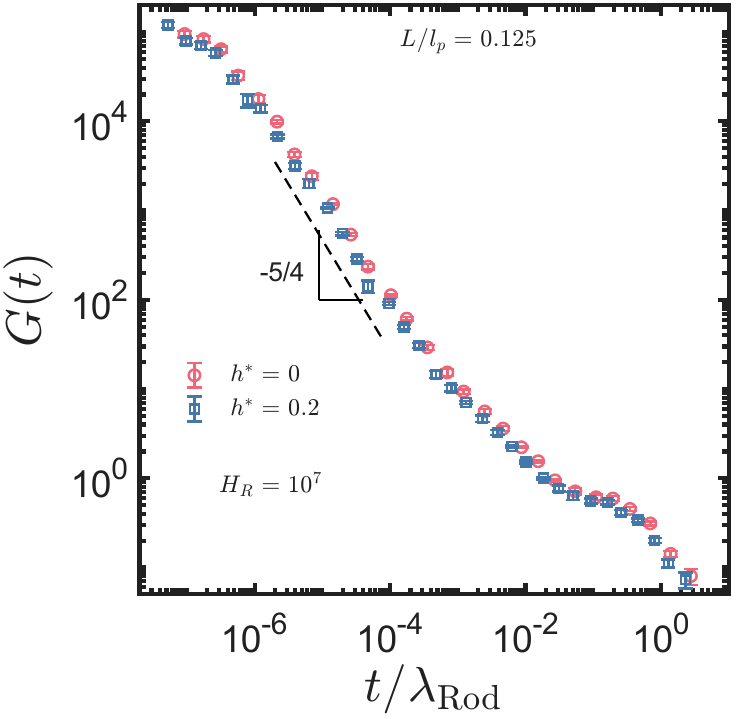} \\
  (e) & (f) \\
\end{tabular}
\caption{The nondimensional relaxation modulus $G(t)$ of chains with varying semi-flexibilities ranging from, (a) Flexible, (b) \(L / l_{p} = 50\), (c) \(L / l_{p} = 10\), (d) \(L / l_{p} = 4\), (e) \(L / l_{p} = 1\) and (f) \(L / l_{p} = 0.125\) have been shown without hydrodynamic interactions (circles) and with hydrodynamic interactions (squares) for \(N_{b}\) = 32. Here the slopes represent the intermediate power law scaling for each cases. Initial spring relaxation is not displayed in all the cases.}
\label{fig11}
\end{figure*}

As \(L/l_{p}\) increases, the intermediate $(-5/4)$ regime gradually vanishes, and the predicted short-time scaling of $(-3/4)$ becomes accessible within the simulation window. The results are in excellent agreement with the intermediate power-law scalings previously reported for semiflexible bead–rod chains in the free-draining case by \citet{Dimitrakopoulos2001}. In their work, the entire intermediate regime of \(G(t)\) was fitted with a single power law, \(G(t) \propto t^{-\alpha}\), with the exponent \(\alpha\) depending sensitively on chain flexibility. For example, for \(L/l_p = 2\), they obtained \(\alpha = 3/4\). The present study extends this analysis over a much broader range of \(L/l_p\) values.

\begin{figure}[t]
\centering
\includegraphics*[width=8.5cm]{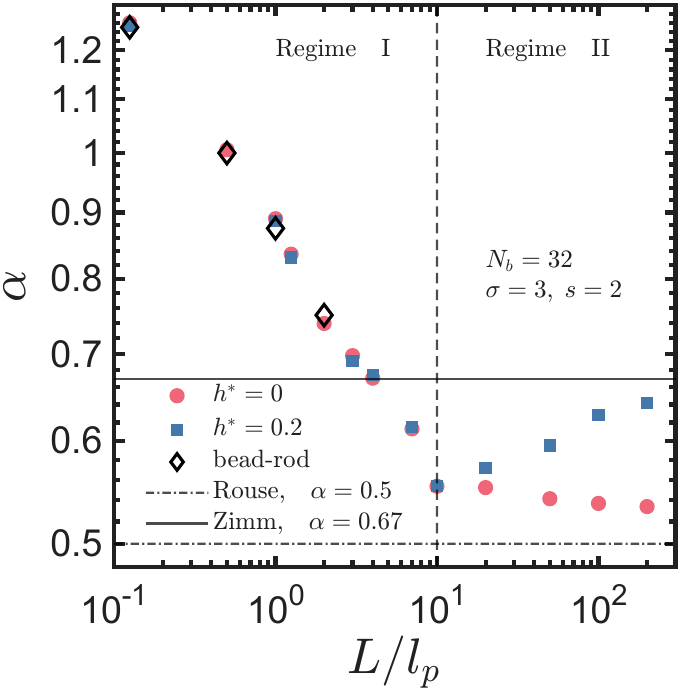}
\caption{Intermediate power-law scaling ($\alpha$) variations with \(L / l_{p}\) reported without hydrodynamic interactions (filled circles) and with hydrodynamic interactions (filled squares). The black diamond symbols are the intermediate power-law scaling reported by \citet{Dimitrakopoulos2001} for semiflexible bead-rod chain results with the free-draining approximation. The solid and dashed lines are the reported slopes Zimm and Rouse theories.}
\label{fig12}
\end{figure}

For small values of \(L/l_{p}\), the relaxation modulus exhibits a distinct plateau at long times, followed by a mono-exponential decay corresponding to the terminal relaxation of a rigid rod of equivalent length. In this regime, the high bending stiffness constrains the chain, causing the intermediate modes of the semiflexible chain to relax almost instantaneously. Once these internal modes have relaxed, the chain dynamics reduce to those of a rigid rod, giving rise to the observed terminal plateau.

As \(L/l_{p}\) increases and the chains become more flexible, they tend to coil rather than remain extended. Consequently, the long-time rodlike orientational relaxation diminishes and eventually vanishes. Instead, the terminal relaxation of flexible chains is governed by the slowest relaxation of their internal modes.

\subsection{\label{sec:hyd_intr} Effect of Hydrodynamic Interactions}

\begin{figure}[t]
\centering
\includegraphics*[width=9cm]{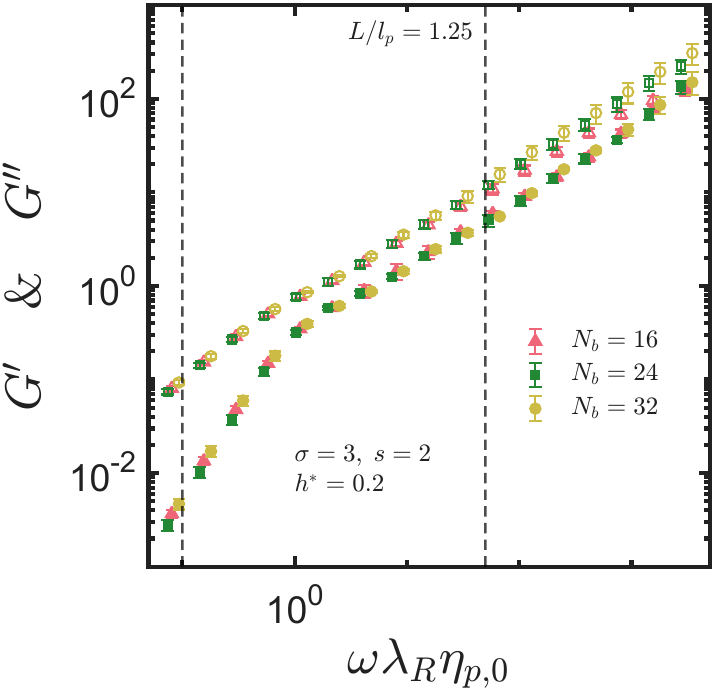}
\caption{Nondimensional dynamic moduli, \({G'}\) (filled symbols) and \({G''}\) (hollow symbols) as a function of frequency. $N_b = 16, 24$ and $32$ bead FENE-Fraenkel chains are used for $L / l_{p} = 1.25$. The dashed lines corresponds to the frequency windows for which the experimental data are available.}
\label{fig13}
\end{figure}

\begin{figure*}[tph]
\centering
\begin{tabular}{cc}
\includegraphics[width=8.5cm]{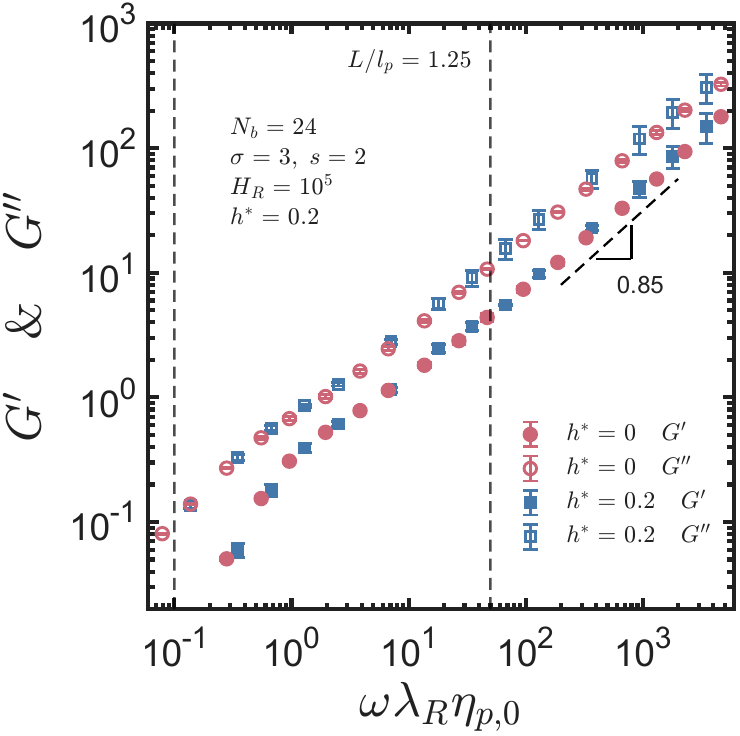} &
\includegraphics[width=8.5cm]{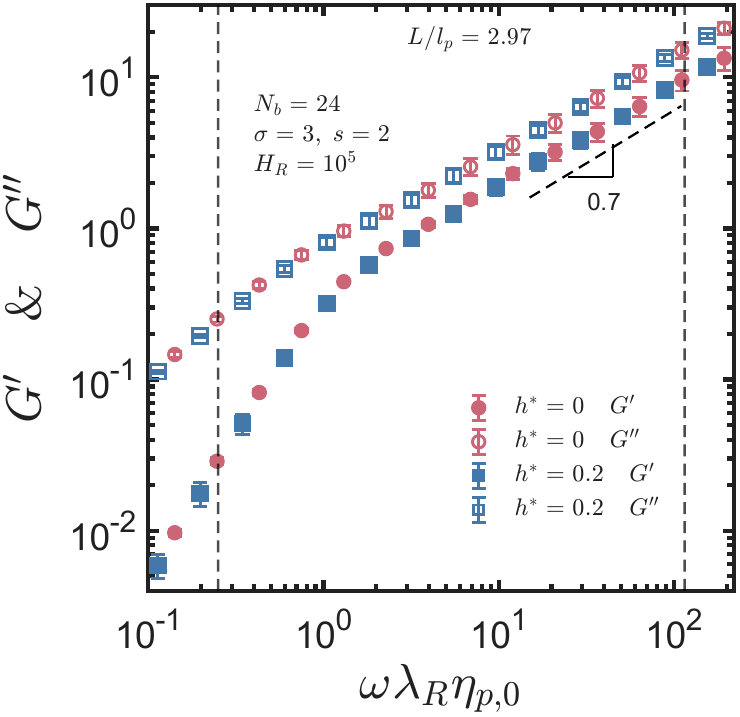} \\
(a) & (b)
\end{tabular}
\caption{Nondimensional dynamic moduli, \({G'}\) (filled symbols) and \({G''}\) (hollow symbols) as a function of frequency for (a) \(L/l_{p} = 1.25\) and (b) \(L/l_{p} = 2.97\), with and without hydrodynamic interactions. The dashed vertical lines corresponds to the frequency windows for which the experimental data are available. Here the slopes represent the intermediate power law scaling for each cases.}
\label{fig14}
\end{figure*}

As shown in the previous sections, the FENE–Fraenkel spring model successfully reproduces the behavior of a bead-rod chain across different semiflexibilities under the free-draining approximation. It captures the key features of semiflexible chains in agreement with existing theoretical and simulation results. To extend these findings, we now examine the effect of hydrodynamic interactions on the linear viscoelastic response.  
It is well established that the Zimm theory demonstrates the importance of hydrodynamic interactions in accurately capturing the dynamics of flexible polymers in dilute solution. In particular, the Rouse and Zimm theories predict distinct intermediate-time power-law scalings in the linear viscoelastic response. However, to the best of our knowledge, the role of fluctuating hydrodynamic interactions in determining the linear viscoelastic behavior of dilute semiflexible polymers has not been systematically explored. Here, we address this gap by investigating the influence of hydrodynamic interactions on the relaxation modulus $G(t)$. We note that while the Zimm model incorporates hydrodynamic interactions in a pre-averaged analytical form, Brownian dynamics simulations are an exact numerical solutions of the governing equations, and as a result, fluctuations are taken into account.

Fig.~\ref{fig11} presents the relaxation modulus \(G(t)\) for a FENE-Fraenkel chain with \(N_b = 32\), and the maximum $H_R$ required to obtain the bead-rod plateau. The chain rigidity is varied from the fully flexible case (Fig.~\ref{fig11}~(a)) to $L/l_p = 0.125$ (Fig.~\ref{fig11}~(f)). For a fully flexible chain, the inclusion of hydrodynamic interactions produces an intermediate slope of $(-2/3)$, consistent with Zimm theory, whereas the free-draining case exhibits the Rouse prediction of $(-1/2)$. As the chain becomes stiffer, the effect of hydrodynamic interactions diminishes, and the difference in intermediate slopes decreases, eventually disappearing for $L/l_p = 10$.  

This trend is quantified in Fig.~\ref{fig12}, which shows the intermediate-time scaling exponent $\alpha$ as a function of $L/l_p$. Two regimes can be distinguished. In Regime I (small $L/l_p$), no difference is observed, indicating that the influence of hydrodynamic interactions is negligible. In Regime II (large $L/l_p$, approaching the flexible limit), there is a deviation of $\alpha$ from the free-draining case, reflecting the relevance and importance of hydrodynamic interactions. The crossover between these regimes occurs around $L/l_p = 10$, beyond which chains remain semiflexible but require hydrodynamic interactions to be accurately captured.

The entire discussion in this section is relevant only to systems in the limit of infinite dilution, where hydrodynamic interactions are present on all length scales and no hydrodynamic screening exists. However, in solutions of finite concentration, where multiple chains interact hydrodynamically, the physics becomes considerably richer as concentration increases. As is well known, beyond a certain critical concentration \(c/c^*\), hydrodynamic interactions become screened. Furthermore, in systems that are crosslinked or entangled, multiple additional length scales emerge, such as the mesh size \(\xi\), the distance between crosslinks \(l_c\), and the entanglement length \(l_e\). The relative magnitudes of these length scales with respect to the persistence length \(l_p\) are anticipated to be important in the discussion of the length scales over which hydrodynamic interactions are screened. None of these effects are addressed in the present work, which is restricted to the single-chain (infinite-dilution) regime. Nevertheless, the framework developed here establishes the foundation for such studies, by identifying the key parameters required in bead–spring chain simulations to accurately reproduce the viscoelastic response of semiflexible systems. This methodology can now be extended to investigate solutions at finite concentrations, where hydrodynamic screening, entanglement effects, and crosslinking can be systematically incorporated.

\begin{figure}[tbph]
\centerline{
\begin{tabular}{c}
\includegraphics[width=7cm]{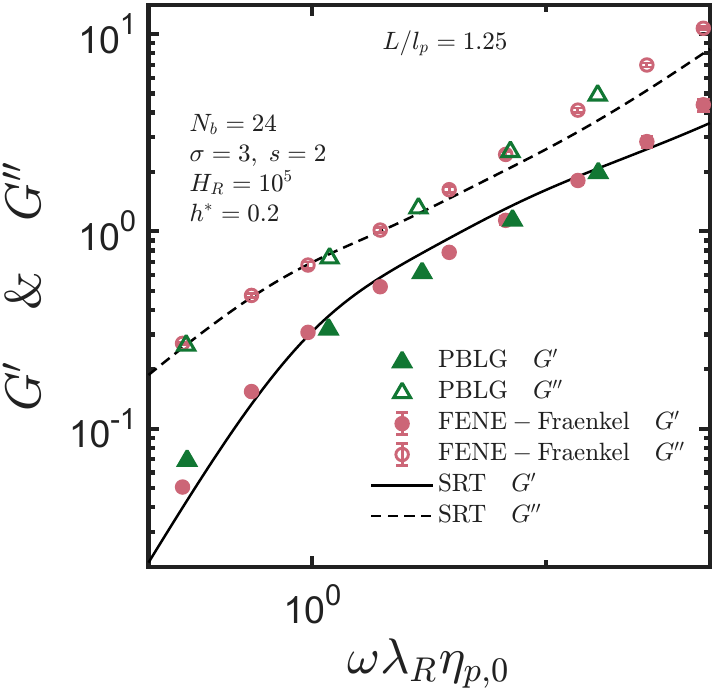} \\
(a) \\[6pt]
\includegraphics[width=7cm]{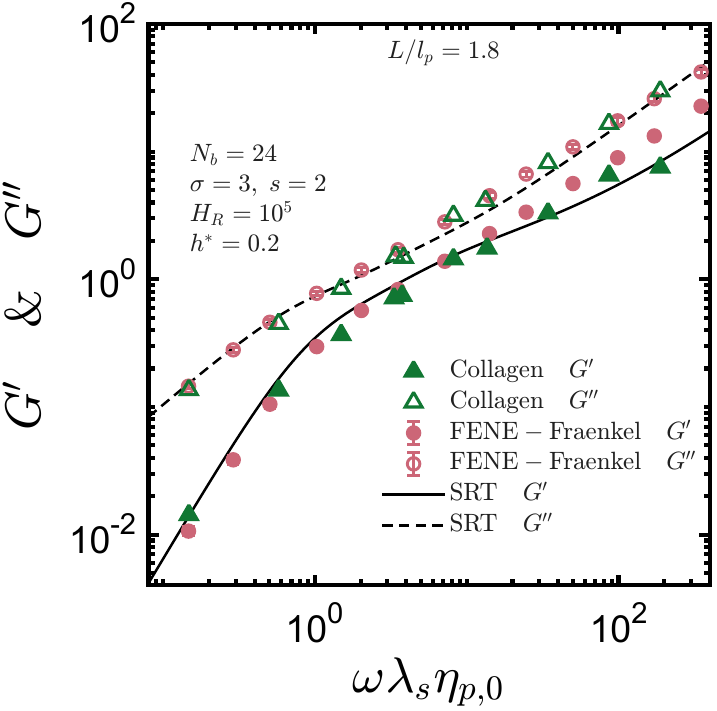} \\ 
(b) \\[6pt]
\includegraphics[width=7cm]{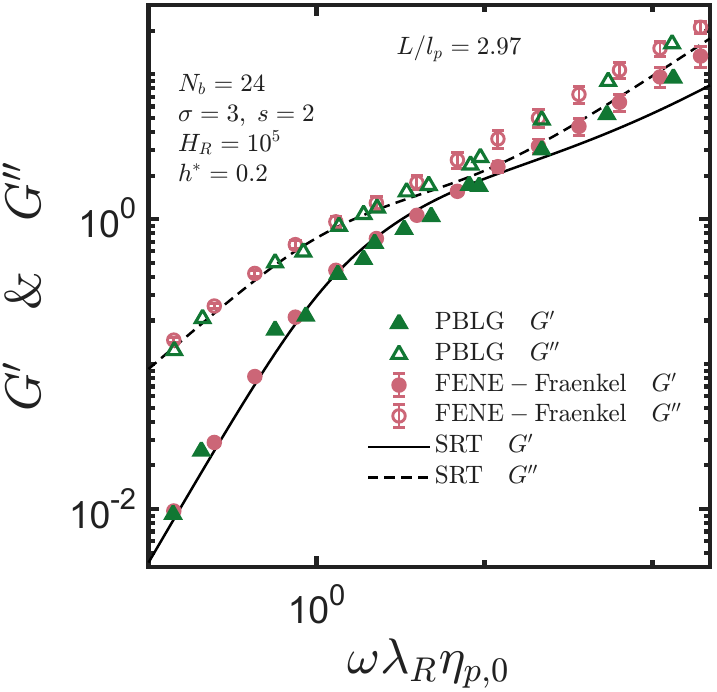} \\ 
(c) \\
\end{tabular}
}
\vspace{-5pt}
\caption{Comparison with experimental data for (a) PBLG, $L/l_p = 1.25$ (b) Collagen, $L/l_p = 1.8$ (c) PBLG, $L/l_p = 2.97$. The data for PBLG polymers in m-cresol solvent is taken from \citet{Warren1973} and Collagen is taken from \citet{Nestler1983} at infinitely dilute concentration. The frequency is scaled with zero shear rate viscosity.}
\label{fig15}
\end{figure}

\subsection{\label{sec:exp} Comparison with experiments}

In the previous sections, the linear viscoelastic response of dilute semiflexible polymer solutions was examined across the full range of \(L / l_{p}\) values using the FENE–Fraenkel bead-spring chain model. These simulations effectively capture the viscoelastic response in both the presence and absence of hydrodynamic interactions, providing a unified framework that spans the range of stiffness from rigid rods to flexible polymers. In this section, the model is validated by direct comparison with existing experimental data for biopolymer systems. Existing theoretical models are also included alongside the simulation results to illustrate how the present approach improves on prior results.

The dynamic moduli predicted by the SRT theory were previously compared by \citet{Shankar2002} with the experimental data of poly-\(\gamma\)-benzyl-L-glutamate (PBLG) in m-cresol solvent, reported by \citet{Warren1973}. It should be noted that the experimental data reported here correspond to dynamic moduli obtained for different polymer concentrations and subsequently extrapolated to the infinite-dilution limit, where inter-chain interactions are negligible \citep{Warren1973}. While the SRT model provided good agreement for chain lengths of \(L = 108 \,\mathrm{nm}\) and \(L = 162 \,\mathrm{nm}\), it was not extended to the longest chain of \(L = 367 \,\mathrm{nm}\), since the theory is strictly valid in the limit \(L / l_{p} \ll 1\). In their analysis, the persistence length of PBLG was estimated to be \(l_p \approx 130 \,\mathrm{nm}\).

To validate the present model, simulations were carried out at the same \(L / l_{p}\) values as in these experiments, considering both the free-draining and hydrodynamic interaction cases. The dynamic moduli were obtained from curve fits to the relaxation modulus, as described in Sec.~\ref{sec:properties}. The frequency axis was normalised using the nondimensional zero shear rate viscosity \(\eta_{p,0}\) (which is identical to the longest nondimensional relaxation time defined in terms of the intrinsic viscosity), consistent with experimental literature, to enable direct comparison between simulation and experimental data across different polymer systems. 

Another key factor in the simulations is chain discretization, determined by the number of beads \(N_b\) in the bead–spring chain. The choice of \(N_b\) directly influences accuracy, particularly at intermediate timescales, as discussed earlier. While increasing \(N_b\) improves resolution, it also increases computational cost. To balance accuracy and efficiency, three discretizations (\(N_b = 16, 24, 32\)) were tested for \(L / l_{p} = 1.25\), as displayed in Fig.~\ref{fig13}. The results show that \(N_b = 24\) is sufficient, as the dynamic moduli from \(N_b = 24\) and \(N_b = 32\) overlap across the relevant frequency range. Therefore, all results reported henceforth use \(N_b = 24\).

Fig.~\ref{fig14} compares the dynamic moduli obtained from simulations with and without hydrodynamic interactions for two values of \(L / l_{p}\). Since the representative \(L / l_{p}\) values fall within Regime II of Fig.~\ref{fig12}, the moduli with and without hydrodynamic interactions overlap when scaled with the zero shear rate viscosity across the relevant frequency window. Based on this observation, only the results including hydrodynamic interactions are used for comparison with experimental data.

The simulations were also compared with experimental data for collagen biopolymers reported by \citet{Nestler1983}, where polymers had an average contour length of \(\sim 300 \,\mathrm{nm}\) and a persistence ratio of \(L/l_p \approx 1.8\). The experiments were performed in two solvents: Solvent C (0.3 M acetate buffer, pH 4.0, with 6 mM NaCl) and Solvent G (identical composition with 70\% glycerol by weight). Fig.~\ref{fig15} compares the predictions of the current simulations with PBLG and collagen experimental data. The bead-spring chain results show excellent agreement across the full frequency range, whereas the SRT theory begins to deviate at higher frequencies, with the discrepancy worsening as \(L/l_p\) increases. In contrast, the simulations reproduce both low- and high-frequency behavior, demonstrating the robustness of the FENE–Fraenkel bead–spring chain model in replicating the linear viscoelastic response of semiflexible polymers.

\section{\label{sec:conclusion}Conclusions}
In this study, a mesoscopic model was successfully developed to investigate the linear viscoelastic response of a single semiflexible polymer chain in the  limit of infinite dilution. Brownian dynamics simulation was carried out for a coarse-grained bead-spring chain with hydrodynamic interactions incorporated to compute the linear viscoelastic properties. The model employs a versatile FENE-Fraenkel spring force law and a bending potential to control chain stiffness. The implementation of hydrodynamic interactions for a semiflexible chain has been validated by comparison with the dynamic properties predicted using an MPCD-MD simulation. The linear viscoelastic behavior of a semiflexible chain under the free-draining approximation has been validated with analytical theories for an inextensible semiflexible rod and semiflexible bead-rod simulation results. For the first time, the effect of hydrodynamic interactions on the linear viscoelasticity of semiflexible chains has been comprehensively evaluated.

\begin{enumerate}
  
\item In the absence of hydrodynamic interactions, the simulation data for a chain of a given $L/l_p$ with varying $H_R$ and $N_b$ collapses onto a master curve, which overlaps with the bead-rod simulation results and the SRT \citep{Shankar2002} for the given $L/l_p$. Once the spring relaxations have occurred, the FENE-Fraenkel bead-spring chain accurately replicates the linear viscoelastic behavior of a semiflexible bead-rod chain. The time frame for this bead-rod behavior can be extended by either increasing the spring stiffness or increasing the discretization of the chain.

\item The stress relaxation modulus was shown to exhibit a distinct power-law behavior at intermediate times. The exponent of this power law varies systematically with chain stiffness, ranging from $(-1/2)$ for a flexible chain to $(-5/4)$ for a  stiff semiflexible chain. The initial $(-3/4)$ predicted by theory was not accessible within the current simulation time window for a stiff semiflexible chain, but it could be captured at later times as $L/l_p$ increases.

\item Hydrodynamic interactions were observed to be negligible for stiff semiflexible chains with $L/l_p$ values up to $L/l_p=10$. Hydrodynamic interactions begin to affect the linear viscoelastic response once the chains become more flexible in nature. In the limit of a fully flexible chain, the intermediate power-law slope approaches the Rouse-like scaling $(-1/2)$ for free draining chains and Zimm-like scaling $(-2/3)$ for chain with hydrodynamic interactions. However, chains with $L/lp>10$, which show distinct behavior with hydrodynamic interactions are still semiflexible in nature. This study concludes that even for a semiflexible chains it is crucial to consider the effect of hydrodynamic interactions for a particular range of $L/l_p$.

\item Comparison with the experimental data shows that a semiflexible FENE-Fraenkel chain is also quantitatively able to reproduce the linear viscoelastic behavior of various semiflexible polymer systems. A bead-spring chain with $N_b = 24$ and $H_R=1 \times 10^{5}$ is accurately able to reproduce the dynamic moduli for the experimental data for the entire frequency range, particularly at higher frequencies where the SRT shows a deviation from the experimental data.

\end{enumerate}

This work provides a robust and computationally tractable framework for understanding the linear viscoelastic properties of semiflexible polymers. By bridging the gap between computationally expensive bead-rod models and bead-spring chains that lack inextensibility, the model provides a useful approach for future research.

The present classification of semiflexible and rod-like regimes, based on the ratio $L/l_p$, is valid for isolated single chains in the infinite-dilution limit which is investigated here. In solutions or crosslinked systems at finite concentrations, additional characteristic length scales such as the mesh size $\xi$, entanglement length $l_e$, and crosslink length $l_c$ become increasingly important in determining the viscoelastic response. Future work will focus on extending this model to explore the viscoelastic properties of such concentrated and networked systems, where hydrodynamic screening, inter-chain correlations, and connectivity effects become significant.

\section*{Author contributions}
AV performed the computer simulations, collected the data and contributed data and analysis tools; AV, PS and RPJ conceived and designed the analysis, performed the analysis, and wrote the paper.

\section*{Conflicts of interest}
There are no conflicts to declare.


\section*{Acknowledgements}
This research was undertaken with the assistance of resources from the National Computational Infrastructure (NCI Australia), an NCRIS enabled capability supported by the Australian Government. This work was also employed computational facilities provided by Monash University through the DUG, MASSIVE and MonARCH systems. We also acknowledge the funding and general support received from the IITB-Monash Research Academy.







\providecommand*{\mcitethebibliography}{\thebibliography}
\csname @ifundefined\endcsname{endmcitethebibliography}
{\let\endmcitethebibliography\endthebibliography}{}

\end{document}